\documentclass[journal,12pt,onecolumn,draftclsnofoot]{IEEEtran}

\usepackage[utf8]{inputenc} 
\usepackage[T1]{fontenc}
\usepackage{url,hyperref}
\usepackage{ifthen}
\usepackage{bbm}
\usepackage[noadjust]{cite}
\usepackage[cmex10]{amsmath} % Use the [cmex10] option to ensure complicance
\usepackage [english]{babel}
\usepackage [autostyle, english = american]{csquotes}
\MakeOuterQuote{"}
\usepackage{graphicx}
\usepackage{subcaption}
\usepackage{algpseudocode}
\usepackage{algorithm}

\usepackage{xparse,amssymb,mathtools,color,graphicx,algpseudocode,algorithm,comment,dsfont,amssymb,stackengine,bm}

\newtheorem{theorem}{Theorem}
\newtheorem{lemma}{Lemma}

\newtheorem{corollary}{Corollary}

\newtheorem{remark}{Remark}

\newcommand{\Z}{\mathbb{Z}}
\newcommand{\F}{\mathbb{F}}
\newcommand{\R}{\mathbb{R}}

\newcommand{\cR}{{\cal R}}
\newcommand{\cV}{{\cal V}}
\newcommand{\cC}{{\cal C}}
\newcommand{\cS}{{\cal S}}
\newcommand{\cE}{{\cal E}}
\newcommand{\n}[1]{{\|\mathbf{#1}\|}}
\newcommand{\FMOR}[1]{\textbf{\textcolor{cyan}{*** FIXME ORI: #1 ***}}}

\NewDocumentCommand\sqn{mg}{%
    \|\mathbf{#1}_{\IfNoValueTF{#2}{}{#2}}\|^2%
}

\newcommand{\RomanNumeralCaps}[1]
    {\MakeUppercase{\romannumeral #1}}

\def\b{\mathbf}

\DeclarePairedDelimiter\floor{\lfloor}{\rfloor}
\graphicspath{{../Figures/}} % Specifies the directory where pictures are stored
\DeclareGraphicsExtensions{.pdf,.jpeg,.png,.jpg}

\newcommand\Algphase[1]{%
\vspace*{-.7\baselineskip}\Statex\hspace*{\dimexpr-\algorithmicindent-2pt\relax}\rule{0.45\textwidth}{0.4pt}%
\Statex\hspace*{-\algorithmicindent}\textbf{#1}%
\vspace*{-.7\baselineskip}\Statex\hspace*{\dimexpr-\algorithmicindent-2pt\relax}\rule{0.45\textwidth}{0.4pt}%
}
\DeclareMathOperator{\EX}{\mathbb{E}}% expected value

\begin{document}
\title{Private Information Retrieval Over Gaussian MAC}

\author{\IEEEauthorblockN{Ori Shmuel, Asaf Cohen \thanks{This research was partially supported by MAFAT. Parts of this work will appear at the 2020 IEEE International Symposium on Information Theory (ISIT). The authors are from Ben-Gurion University of the Negev, Israel.
Email: \{shmuelor,coasaf\}@bgu.ac.il }}
%\IEEEauthorblockA{\\Ben-Gurion University of the Negev, \{shmuelor, coasaf, gurewitz\}@bgu.ac.il}
}

\maketitle

\begin{abstract}
Consider the problem of Private Information Retrieval (PIR), where a user wishes to retrieve a single message from $N$ non-communicating and non-colluding databases (servers). All servers store the same set of $M$ messages and they respond to the user through a block fading Gaussian Multiple Access Channel (MAC). The goal in this setting is to keep the index of the required message private from the servers while minimizing the overall communication overhead.

This work provides joint privacy and channel coding retrieval schemes for the Gaussian MAC with and without fading. The schemes exploit the linearity of the channel while using the Compute and Forward (CF) coding scheme. Consequently, single-user encoding and decoding are performed to retrieve the private message. In the case of a channel without fading, the achievable retrieval rate is shown to outperform a separation-based scheme, in which the retrieval and the channel coding are designed separately. Moreover, this rate is asymptotically optimal as the SNR grows, and are up to a constant gap of $2$ bits per channel use from the channel capacity without privacy constraints, for all SNR values. When the channel suffers from fading, the asymmetry between the servers' channels forces a more complicated solution, which involves a hard optimization problem. Nevertheless, we provide coding scheme and lower bounds on the expected achievable retrieval rate which are shown to have the same scaling laws as the channel capacity, both in the number of servers and the SNR. 
\end{abstract}


\section{Introduction}

\IEEEPARstart{T}{he} ability to provide privacy and protection to sensitive data has become a requirement in communication systems nowadays. While cryptography and physical layer security provide various solutions against adversaries which are located outside the system, in some applications privacy is required even from the system's administrators which have access to the data even before transmission. 
In the most basic setting of PIR, which was first introduced by Chor \emph{et al.} \cite{chor1995private}, there are $N$ identical and non-communicating databases (servers) where each stores the same $M$ messages. A user, who is interested in a single message yet wishes to keep the servers ignorant about the identity of that message, generates a series of queries to the servers, which answer them truthfully. His goal is to minimize the overhead needed to attain privacy. The problem of PIR was considered by the Computer Science community extensively, e.g. \cite{gasarch2004survey, ostrovsky2007survey, yekhanin2010private}. Recently, the problem was considered also by the Information Theory community, which gave it a slightly different interpretation, in an effort to characterize the fundamental limits of the problem. Specifically, in the classic PIR problem, the performance metric, referred to as "communication complexity", is the sum of the total upload cost (the size of the queries) and the total download cost (the size of the servers' answers). In the information theoretic formulation, the size of the messages is assumed to be arbitrarily large and thus one may neglect the upload cost. The performance metric in this case is the rate of the PIR scheme, defined as the ratio between the size of the desired message and the total download\footnote{Note that if the communication channel is noisy, the dowloaded information, specifically, the \emph{transmitted data}, contains additional redundancy which must be considered also.}, arriving the user \cite{sun2017capacity}.

Under such a formulation, the PIR capacity, which is the supremum of PIR rates over all achievable retrieval schemes, was presented in \cite{sun2017capacity} for the classical PIR problem. Specifically, \cite{sun2017capacity} showed that the PIR capacity is $C_{PIR}=\left(1+\frac{1}{N}+\frac{1}{N^2}+...+\frac{1}{N^{M-1}}\right)^{-1}=\left(1-\frac{1}{N}\right)/\left(1-\left(\frac{1}{N}\right)^M\right)$ and provided an achievable retrieval scheme. 

Naturally, many extensions for the PIR problem were considered. For example, robust PIR with colluding servers was considered in \cite{sun2017capacityColluding, tajeddine2017private} where some of the servers may exchange the queries submitted between them. An extension to Byzantine servers, which respond with erroneous answers, where the errors may be unintentional or even deliberate can be found in \cite{banawan2018capacityByzantine}. In \cite{sun2018capacity}, the case of symmetric PIR was investigated, where the user learns nothing on the other unwanted messages. In \cite{sun2017optimal}, the minimum download cost for arbitrary message size $L$ was investigated. In \cite{tian2019capacity}, using a new PIR code construction, the optimal message size and upload cost was presented. Extensions involving an eavesdropper, i.e., secure PIR, can be found in \cite{yang2018private,banawan2020private}. PIR with side information was examined in \cite{wei2019capacity}, where an additional prefetching phase to the user cache is possible. This phase enables the servers to have partial knowledge on the side information the user has. The above mentioned works assume that the content on the servers is the same (i.e., a repetition code), which on one hand provides the highest resistance against errors but on the other requires extremely large storage cost. Thus, recent works also considered the PIR problem for coded servers, which offers the same amount of data reliability with overall less storage cost \cite{shah2014one, chan2015private, tajeddine2018private, banawan2018capacity,zhu2019new,tajeddine2019private}. Interestingly, the capacity of the PIR for coded servers was found to be a function of the coding rate $R_{code}$ and the number of messages $M$ \cite{ banawan2018capacity}. Specifically,  $C_{PIR}=\left(1-R_{code}\right)/\left(1-\left(R_{code}\right)^M\right)$, where one can observe that the case of repetition coding, i.e., $R_{code}=\frac{1}{N}$, assumed in \cite{sun2017capacity} comes as a special case.

In both the classical PIR problem, as well as the extensions mentioned above, it is assumed that the servers answer the user through noiseless orthogonal channels (bit-pipes), which means that the user receives $N$ separate responses, from which it needs to decode the desired message. However, in many practical scenarios, the communication channel endures some kind of noise. For example, random packets are being dropped due to congestion or may be corrupted in some way (e.g., wireless channels). In \cite{banawan2019noisy}, the PIR problem with noisy orthogonal links was investigated. Therein, the user observes a noisy version of the servers' responses, which may endure asymmetric traffic constrains. \cite{banawan2019noisy} provided upper and lower bounds on the retrieval rate and showed that the channel coding and the retrieval scheme are almost separable, in the sense that both must agree in advanced on the capacities of the channels, yet given these capacities, the schemes can be designed separatly. In addition, they considered a variant of the PIR problem for which the responses of the servers are mixed before reaching the user. Such a variant may represent a Multiple Access Channel (MAC)\footnote{An example of such a scenario may be when a user is trying to retrieve privately a file from several wireless base-stations.}.

In the MAC-PIR problem considered in \cite{banawan2019noisy}, a binary additive MAC and logical conjunction/disjunction MAC were investigated. In this case, as opposed to noisy PIR with noisy orthogonal links \cite{banawan2019noisy}, the channel coding and the retrieval schemes should be designed together. Specifically, the authors provided schemes that can achieve the full channel capacity while still being private. This is done by using the linearity of the MAC and the ability to compute a function of the transmitted servers' responses. The capacity for the binary additive MAC model and the limits, in general, for computations over MAC were given in \cite{nazer2007computation}. Thus, \cite{banawan2019noisy} further enlightens us regrading the channel's computational capabilities, now working in favour of privacy in the PIR problem.
%\FMOR{In the long version add here a sentence with cites that the problem of separation was engaged in multi terminal communication problems in the past.}

%PIR capacity by using the physical properties of the MAC channel. That is, the user uses the linear property of the channel and eventually tries to compute a function of the transmitted servers' responses. 
%\textcolor{blue}{\cite{banawan2019noisy} essentially shows that the channel capacity is achieved while still being private. The capacity for the binary additive MAC model and the limits for the general problem of computations over MAC was given also in \cite{nazer2007computation}.}
%It turns out that in \cite{nazer2007computation}, the limits for the problem of computations over MAC was considered under a different context. And in fact, a similar channel model as in \cite{banawan2019noisy} was considered with similar results about the capacity of the channel and specifically on the non-separability of the source and channel coding. So, basically the limits for computation over MAC were already considered, however, the work in \cite{banawan2019noisy} was the first to present the channel computation ability in favour the privacy in the PIR problem.

\subsection{Main Contributions}
In this work, we consider the PIR problem for the Gaussian additive MAC with and without fading; these are the most common models for wireless networks. Although the optimal encoding and decoding schemes for the MAC are quite clear, adding the privacy constraints that this problem imposes leads to new challenges for which these optimal schemes provide insufficient performance. That is, separating between the channel coding and the PIR coding for the Gaussian MAC is sub-optimal. We start by providing a basis for comparison by considering an achievable PIR scheme which is based on separation. We then provide a joint coding scheme based on lattice codes, along with analysis on its achievable rate, which exploits the additive nature of the channel as well as the linear properties and structure of lattice coding. To the best of our knowledge, this is the first work which combines lattice coding with PIR. For non-fading channels, we show that such a joint scheme outperforms separation as the number of servers and SNR grows. Specifically, for large number of servers, the joint scheme performance is twice as good as separation. Moreover, we show that the achievable PIR rate is within a constant gap ($2$ bits per channel use) from the capacity of the channel \emph{without privacy constraints}, which is used as a global upper bound on the performance in order to assess the tightness of our results. Furthermore, in the limit of high SNR, this achievable PIR rate approaches the capacity. That is, the scheme is asymptotically optimal. Thus, \emph{privacy can be achieved with negligible loss}. In addition, the suggested coding scheme provides simplicity in attaining privacy compared to known PIR schemes.

The extension to the case of fading channels complicates the problem since the fading imposes asymmetric links between the servers and the user. In general, asymmetry fundamentally hurts the retrieval rate \cite{banawan2019asymmetry}. However, we provide a PIR scheme with analysis on the expected achievable retrieval rate and show that it overcomes this asymmetry by smart aggregation of servers and the use of lattice coding. Another important issue is the availability of Channel State Information (CSI) to the servers. Obviously, when the CSI is globally known, the servers can improve their transmission rate regardless of the specific coding scheme. On the other hand, prior to their transmission, the user may convey the CSI or part of it in the queries. We thus compare our results with the capacity of the channel with and without CSI at the Transmitter (CSIT). We show that even if the CSIT is not available, and the user does not explicitly send the channel coefficients, the PIR scheme implicitly generates cooperation and information between the servers through the queries, as if CSIT exists. 

As mentioned, our suggested PIR scheme aggregates servers together to cope with the asymmetry. Yet, the aggregation which maximize the rate is hard to obtain due a complex optimization problem. Nevertheless, we provide two lower bounds on the expected PIR rate for two different (sub-optimal) ways of servers aggregation. These lower bounds are shown to achieve the scaling laws of the channel capacity without CSIT either with $N$ or with the transmission power $P$. In addition, we provide numerical results, based on a heuristic algorithm for servers aggregation, which show that the PIR rate is not only higher than the capacity without CSIT, but is within a constant gap from the capacity with CSIT.

\subsection{Paper Outline}
The paper is organized as follows. In Section \ref{Sec-Model and Problem Statement}, the system model and the problem statement are described. In Section \ref{sec-PIR for AWGN MAC}, we present our suggested PIR scheme for non-fading AWGN MAC. We compare this scheme to an achievable PIR scheme by separation and to the capacity of the channel without privacy constraints. Finally, Section \ref{sec-Joint PIR scheme for the block-fading AWGN - MAC} extends our suggested PIR scheme to the case of a block-fading channel.

\section{System Model and Problem Statement}\label{Sec-Model and Problem Statement}

\subsection{Notational Conventions}
Throughout the paper, we will use boldface lowercase to refer to vectors, e.g., $\b{h} \in \R^L$, and boldface uppercase to refer to matrices, e.g., $\b{H} \in \R^{M \times L}$.  For a vector $\b{h}$, we write $\n{h}$ for its Euclidean norm, i.e. $\n{h} \triangleq \sqrt{\sum_{i}h_i^2}$. We denote by $\b{e}_i$ the unit vector with $1$ at the $i$th entry and zero elsewhere. We assume that the $\log$ operation is with respect to base 2.

\begin{figure}[t]
\centering
    \includegraphics[width=0.35\textwidth]{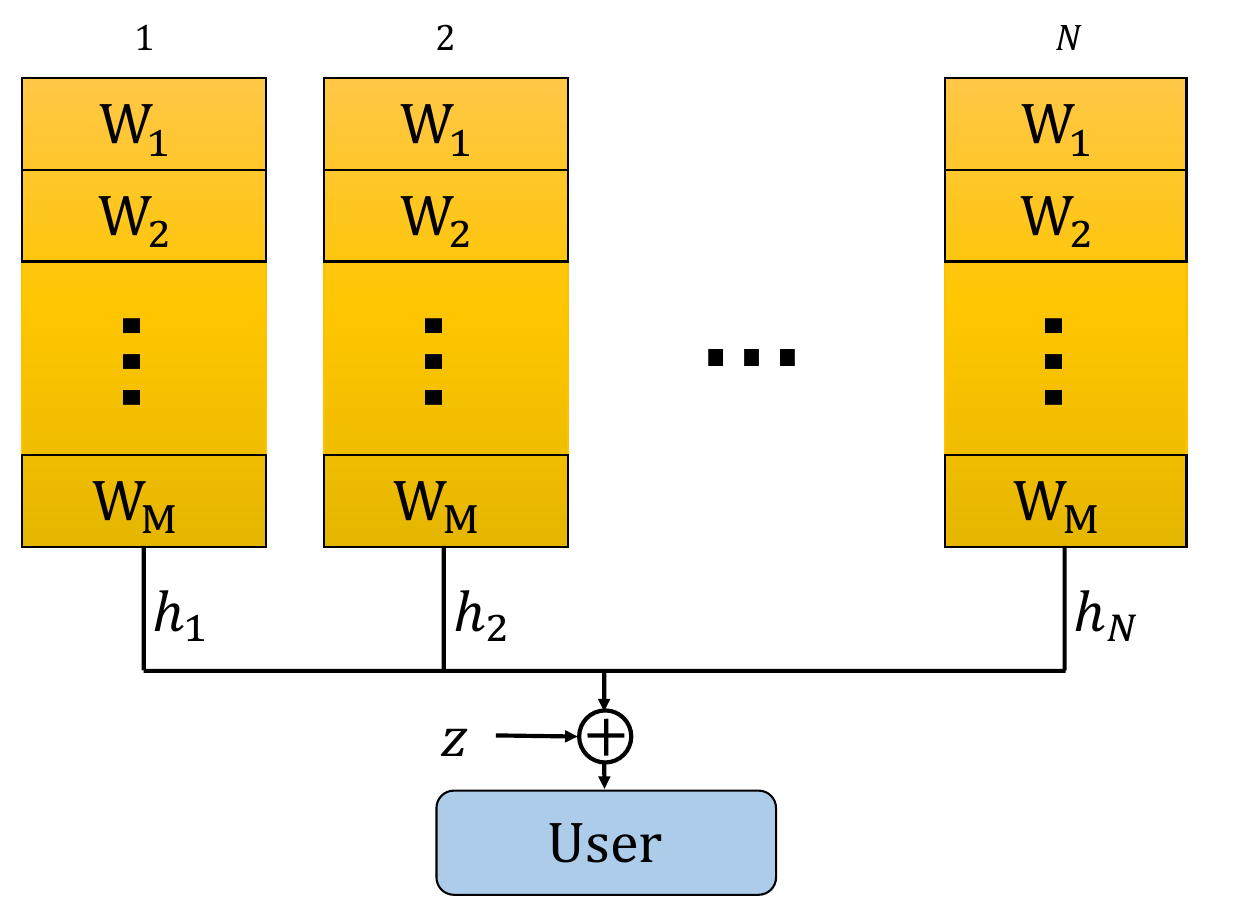}
\caption{System model of $N$ servers connected to a user via a Gaussian MAC.}
\label{fig-PIR_System_model}
\end{figure}

\subsection{System Model}\label{sec-system model}
Consider the basic setting of the PIR problem with $N$ identical and non-communicating servers. Each server stores a set of messages $W_1^M=\{W_1,W_2,...,W_M\}$ of size $L$ each. The messages were drawn uniformly and independently from $\mathbb{F}_p^L$ where $p$ is assumed to be prime\footnote{The assumption that the messages' alphabet is a prime-size finite field generalizes the assumption of binary messages of many PIR works, e.g. \cite{sun2017capacity,sun2017capacityColluding,banawan2019noisy}. In this work, it is a requirement since nested lattice codes are used in the suggested PIR scheme. The construction of such codes require that the original messages' alphabet size $p$ grows like $O(n\log{n})$. Hence, messages over small alphabet size should be mapped to a higher alphabet size.}, i.e., in bit units we have,
\begin{equation}\label{equ-Entropy of messages}
\begin{aligned}
&H(W_l)=L\log{p} \ \text{for} \  l=1,...,M,\\
&H(W_1^M)=ML\log{p}.
\end{aligned}
\end{equation}
In PIR, a user wishes to privately retrieve the message $W_i$, where the index $i$ is assumed to be uniformly distributed on $[1,...,M]$, i.e., $i$ is a realization of $\theta \sim U[1,..M]$, while keeping $\theta$ secret from each server \footnote{We note that the user actually asks for the content of the message located in the $i$th place at the server. That is, we assume that there are $p^L$ possible messages and the number $M$ of messages each server holds may be much smaller.}. Throughout, we use the random variable $\theta$ to denote the (uniformly distributed) message, e.g., when proving privacy, and use the index $i$ to denote its realization. Hence, $Q_j(\theta)$ will be used to stress out that the query depends on the random message required, while $Q_j(i)$ will be used to show the query's dependence on the specific index $i$. Accordingly, the user generates a set of $N$ queries $Q_1(i),Q_2(i),...,Q_N(i)$, one for each server, which are statistically independent with the messages (as those are not known to him). 
That is, we have
\begin{equation*}
I(W_1^M;Q_1(\theta),...,Q_N(\theta))=0.
\end{equation*}
The $k$th server responds to its query with a message (or codeword) $\b{x}_k(i)$ of size $n$. This answer is a deterministic function of the messages and the query. That is, for all $k$ we have, 
\begin{equation*}
H(\b{x}_k(\theta) | W_1^M,Q_k(\theta))=0.
\end{equation*}
Also, to ensure privacy, the queries should not reveal the desired index $i$ to the servers. 
%that is, from the perspective of the servers, the queries should be identically distributed between the messages. That is, for all indices  $i,j\in\{1,...,M\}$,
%\begin{equation}\label{equ-queries are the same between messages}
%\left(Q_k(i),\b{x}_k(i),W_1^M \right) \sim \left(Q_k(j),\b{x}_k(j),W_1^M \right).
%\end{equation}
Consequently, this implies that for each server $j$ the index $\theta$ of the desired message is independent of the query and the answer, that is, the privacy constraint is, 
\begin{equation}\label{equ-queries messages and answers are independent of index}
I(\theta;Q_j(\theta),\b{x}_j(\theta),W_1^M)=0 \ \text{for all } j\in\{1,...,N\}.
\end{equation}
We assume the servers receive the queries through independent control channels, and do not have access to each other's queries or answers.

In this work, we consider the problem of PIR over the Gaussian MAC, and the block-fading Gaussian MAC as depicted in Figure \ref{fig-PIR_System_model}. Accordingly, the user observes a noisy linear combination of the transmitted signals from the servers through the channel,
\begin{equation}\label{equ-the channel}
\b{y}(i)=\sum_{k=1}^{N}h_{k}\b{x}_k(i)+\mathbf{z},
\end{equation}
where $h_{k} \sim \mathcal{N}(0,1)$ are the real channel coefficients and $\b{z}$ is an i.i.d., Gaussian noise, $\b{z} \sim \mathcal{N}(0,\b{I}^{n\times n})$. Note that the index $i$ in the received input denotes the private index of the desired message and not time. Let $\b{h}= (h_{1},h_{2},...,h_{N})^T$ denote the vector of channel's coefficients. We assume a memoryless block-fading channel model, i.e., the channel remains constant during the period of codeword transmission of size $n$; we assume that in each slot the user knows the channel vector while the servers do not have this information\footnote{Note that in case the channel coefficients are globally known the PIR rate given in this work can be further improved since the servers may design their codebooks accordingly (see \cite{zhu2016gaussian}). Furthermore, in Theorem \ref{the-retrieval rate for fading MAC} we show that this global knowledge does not affect our privacy scheme.}. When we examine the Gaussian MAC without fading, we fix $h_k=1$ for $k=1,...,N$. In addition, we assume an average power constraint on the codewords, i.e., $\|\b{x}_k\|^2 \leq nP$. 

Upon receiving the mixed response $\b{y}(i)$ from all the servers, the user must be able to decode the required message $W_i$. Let $\widehat{W_i}$ denote the decoded message at the user and define the error probability of decoding a message as
\begin{equation}\label{equ-probability of error definition}
P_e(L)\triangleq P_r(\widehat{W_i} \neq W_i).
\end{equation}
We require that $P_e(L)\rightarrow 0$ as $L$ tends to infinity.

\subsection{Performance Metric}
In the noiseless, orthogonal case, the PIR rate (or retrieval rate) is defined by the total desired bits divided by the total received bits \cite{sun2017capacity,banawan2018capacity}. Specifically, 
\begin{equation}
R_{PIR}\triangleq \frac{H(W_\theta)}{\sum_{k=1}^{N}H(\b{x}_k(\theta))}\triangleq\frac{L\log{p}}{D},
\end{equation}
where $D$ is the total bits dowloaded from the servers' answers. Accordingly, the above retrieval rate definition describes \emph{only the coding rate (or redundancy) which is needed to keep the message private}. 

When assuming a noisy channel, the servers' answers should also be resilient to the channel's errors and the PIR rate should take into account also the redundancy of the channel coding. When the channel coding and the PIR schemes are designed \emph{separately}, such a metric is easy to acquire. In fact, as mentioned in the introduction, the issue of separation between the PIR and the channel coding schemes was addressed in \cite{banawan2019noisy} for the case of asymmetric noisy orthogonal channels (i.e., different capacities for each channel), for which the authors showed that the two coding schemes are almost separable\footnote{Since the capacities of the channels are asymmetric, the schemes must agree in advance on the amount of information each server can send reliably to the user. The explicit coding scheme is not affected.} and thus applying the capacity-achieving channel code is optimal. As a special case, when all channels are symmetric with a certain channel capacity $C$, each server encodes its $d$ symbols into a $n$-length codeword and sends it to the user. The achievable channel transmission rate is thus, $R_c^p=\frac{d}{n}\log{p}$ (bits per channel use). According to Shannon's channel coding theorem \cite{cover2012elements}, there exist a sequence of codes, $\cC_n$, with probability of error $P_e(\cC_n)$ that tends to zero as the codeword length, $n$, grows, as long as $R_c^p<C$. Thus, the PIR rate, which includes the noisy channel rate, can be upper bounded by,
\begin{equation*}
\begin{aligned}
R_{PIR}^s&=\frac{L\log{p}}{Nn}=\frac{L\log{p}}{N \frac{d}{R_c^p}\log{p}}=\frac{L\log{p}}{D}R_c^p<C_{PIR} \cdot C \ \left[\frac{\text{Bits}}{\text{Ch. use}}\right],
\end{aligned}
\end{equation*}
where now $C_{PIR}$ is the PIR capacity as given in \cite{sun2017capacity} for finite field of size $p$. Essentially, the above represents the maximal rate one can achieve when there is separation between the PIR and the channel coding schemes.

In this work, we show that for the AWGN MAC one can gain better performance when the PIR scheme and the channel coding are designed together. We provide a joint privacy-channel coding scheme that uses the additive nature of the channel in the design of the queries and answers, and as a result significantly decreases the loss incurred by the privacy requirement.

We compare this achievable rate to the full channel capacity without any privacy constraint for sending a single message. This comparison shows the redundancy of the suggested joint scheme one has to tolerate to promise privacy. For that purpose, let us refine the subtleties in the model assumptions with respect to the original PIR problem. We assume that the servers cannot cooperate explicitly, yet cooperation \emph{is possible implicitly} by exploiting the user's queries. In addition, we assume that all transmitting servers transmit with a fixed power $P$, i.e., the power constraint is per-server, and power cannot be allocated differently to different servers. That is, we have per-server power constraint with full cooperation. Accordingly, our model matches the Multiple Input Single Output (MISO) channel with per-antenna power constraint, where each transmit antenna has a separate power budget yet can fully cooperate with other antennas \cite{vu2011miso}. The MISO sum capacity with per-antenna power constraint and fixed channel coefficients, which are globally known, is given by \cite{vu2011miso},
\begin{equation}\label{equ-MISO sum capacity with per-antenna power constraint}
C_{SR}^{MISO}=\frac{1}{2}\log\left( 1+ P\left(\sum_{k=1}^N |h_k|\right)^2\right).
\end{equation}
We note that $C_{SR}^{MISO}$ is higher than the MAC sum capacity (which, in this case, is $\frac{1}{2}\log\left( 1+ P\sqn{h}\right)$), since in the latter each transmitter acts independently to transmit its own message.

When the channel coefficients are known only at the receiver, the ergodic MISO sum capacity with per-antenna power constraint, is given by \cite{vu2011miso},
\begin{equation}\label{equ-ergodic MISO sum capacity with per-antenna power constraint}
C_{SR,ergodic}^{MISO}=\EX_h\left[\frac{1}{2}\log\left( 1+ P\sqn{h}\right)\right].
\end{equation}

When discussing upper bounds on the PIR performance, the capacity expression in \eqref{equ-MISO sum capacity with per-antenna power constraint} will be relevant when assuming a non-fading AWGN MAC, while the second expression in \eqref{equ-ergodic MISO sum capacity with per-antenna power constraint} will be relevant in addition to  \eqref{equ-MISO sum capacity with per-antenna power constraint} for fading channels. Note that both \eqref{equ-MISO sum capacity with per-antenna power constraint} and \eqref{equ-ergodic MISO sum capacity with per-antenna power constraint} consider setups without any privacy constraint. In both PIR scenarios which we consider, the achievable rates of the suggested schemes will be shown to be up to a constant gap from the corresponding full channel capacities, that is, \eqref{equ-MISO sum capacity with per-antenna power constraint} or \eqref{equ-ergodic MISO sum capacity with per-antenna power constraint}.

\subsection{Coding Schemes and Lattice Codes}
The seminal work of Erez and Zamir \cite{erez2004achieving} showed that using lattice encoding and decoding, the full capacity of the point to point AWGN channel is achievable. Following this work, several papers, e.g., \cite{erez2005capacity,  nazer2007computation, nam2008capacity, wilson2010joint, nazer2011compute, hong2013compute, zhan2014integer, ordentlich2014approximate, lim2018joint, shmuel2020compute}, considered different channel models with Gaussian noise, all using lattice codes and their structural properties. The most prominent property is the fact that every linear combination of codewords is a codeword itself\footnote{This property is inherited from linear codes in general.}. We now provide a brief background on lattice codes, which will be useful in the remainder of this paper 

\subsubsection{Nested Lattice Codes}
An $n$-dimensional lattice $\Lambda$ is a discrete subgroup of the Euclidean space $\R^n$ with the ordinary vector edition operation. This implies that if $\lambda_1,\lambda_2 \in \Lambda$ then $\lambda_1+\lambda_2 \in \Lambda$. A lattice quantizer is a map $Q_\Lambda : \R^n \rightarrow \Lambda$, that sends a point $\b{x}\in\R^n$ to the nearest lattice point in Euclidean distance, i.e., $Q_\Lambda(\b{x}) =\arg\min_{\lambda \in \Lambda}\|\b{x}-\lambda\|$. The Voronoi region of $\Lambda$, denoted by $\cV$, is the set of all points in $\R^n$ which are quantized to the zero vector, i.e., $\cV(\Lambda)=\{\b{x}:Q_\Lambda(\b{x}) =\b{0} \}$. The modulo-$\Lambda$ operation is defied as the quantization error of $\b{x} \in \R^n$ with respect to the lattice $\Lambda$, i.e., $\b{x} \ \text{mod} \  \Lambda = \b{x} - Q_\Lambda(\b{x})$. The second moment of a lattice $\Lambda$ is defined as
\begin{equation}\label{equ-second_moment_deffinition}
\sigma_\Lambda^2=\frac{1}{n\text{V}(\cV)}\int_{\cV(\Lambda)} \|\b{x}\|^2 d\b{x},
\end{equation}
where $\text{V}(\cV)$ is the volume of the Voronoi region. The normalized second moment of the lattice, is then given by
\begin{equation}\label{equ-normalized_second_moment_deffinition}
G(\Lambda) \triangleq \frac{\sigma_\Lambda^2}{\text{V}(\Lambda)^{2/n}}.
\end{equation}

Lattice codes are the Euclidean space counterpart of linear codes which provide structure to the codebook. Thus, similar to linear codes, a message $W_m$ with length $L$ is encoded to a codeword with length $n$ using a one-to-one function where, in our case, this codeword is a lattice point. The structure of the lattice (i.e. the positions of the points) and the bounding region, which forms the codebook itself, rule the "goodness" of it as a codebook and the ability to achieve the limits of the communication channel\footnote{A lattice is an unbounded set of points. Thus, exploiting lattices for communication problems requires the bounding of the infinite lattice with a finite shaping region, in order to construct a codebook. In \cite{ordentlich2016simple}, it was shown that there is a simple construction for a sequence of lattice codes which achieve the capacity of the AWGN channel. The construction is based on lifting different sub-codes of a linear code to the Euclidean space using Construction A (\cite{zamir2014lattice}) to form a nested lattice code. Additional information on lattices can be found in \cite{zamir2014lattice}}. 

A nested lattice code is a lattice code which its bounding region is the Voronoi region of a sub-lattice. Formally, let $\Lambda_c$ and $\Lambda_f$ be a pair of $n$-dimensional lattices with Voronoi regions $\cV_c$ and $\cV_f$, respectively, such that $\Lambda_c$ is a subset of $\Lambda_f$, i.e., $\Lambda_c \subset \Lambda_f$. The nested lattice code is thus given by, $\cC=\{\Lambda_f \cap \cV_c\}$, and its rate is equal to \cite{zamir2014lattice},
\begin{equation}\label{equ-lattice rate}
R=\frac{1}{n}\log|\cC|=\frac{1}{n}\log|\Lambda_f \cap \cV_c|=\frac{1}{n}\log|p^L|=\frac{L\log{p}}{n}.
\end{equation}

\subsubsection{Compute-and-Forward}

In \cite{nazer2011compute}, the Compute and Forward (CF) coding scheme, which enables receivers to decode "noisy" linear combinations of transmitted messages, was introduced. Specifically, remembering that according to our channel model, the received answers at the user are attenuated by real (and not integer) attenuations, the receiver of the non-integer linear combination seeks a set of integer coefficients, denoted by a vector $\b{a}$, to be as close as possible to the true channel coefficients and to serve as the coefficients for the linear combination of the received messages. 

The CF scheme uses nested lattice codes for the computation of the linear equation of the transmitted messages. That is, after receiving the noisy linear combination, the user selects a scale coefficient $\alpha \in \R$, an integer coefficient vector $\b{a}=(a_{1},a_{2},...,a_{N})^T \in \Z^N$, and attempts to decode the lattice point $\sum_{k=1}^N a_{k}\b{x}_k$ from $\alpha \b{y}$. Formally, the decoder has
\begin{equation}\label{equ-channel output at the decoder}
\begin{aligned}
	\alpha\b{y}&=\sum_{k=1}^{N}\alpha h_{l}\b{x}_k+\alpha \b{z} \\
			     &=\sum_{k=1}^{N}a_{k}\b{x}_k +\sum_{k=1}^{N}(\alpha h_{k}-a_{k})\b{x}_k+\alpha \b{z}.
\end{aligned}
\end{equation}

Due to the lattice algebraic structure, the relay decodes $\sum_{k=1}^N a_{k}\b{x}_k$ as a codeword, while enduring the noise of $\sum_{k=1}^{N}(\alpha h_{k}-a_{k})\b{x}_k+\alpha \b{z}$, namely, the \emph{effective noise}. The rate of the decoded codeword, i.e., the \emph{achievable rate}, defines a rate region for which all servers must comply with to correctly decode the linear combination. The achievable rate and the optimal scale coefficient are given in the following theorems,

\begin{theorem}[{\cite[Theorem 1]{nazer2011compute}}]\label{the-Computation rate}
For real-valued AWGN networks with channel coefficient vectors $\b{h} \in \mathbb{R}^N$ and coefficient vector $\b{a} \in \mathbb{Z}^N$, the following computation rate region is achievable:
\begin{equation}
\cR(\b{h},\b{a})= \max \limits_{\alpha \in \R} \frac{1}{2} \log^+ \left( \frac{P}{\alpha^2+P\|\alpha \b{h}-\b{a}\|^2} \right),
\end{equation}
\end{theorem}
where $\log^+(x)\triangleq \max \{\log(x),0\}$.
\begin{theorem}[{\cite[Theorem 2]{nazer2011compute}}]\label{the-Computation rate with MMSE}
The computation rate given in Theorem \ref{the-Computation rate} is uniquely maximized by choosing $\alpha$ to be the MMSE coefficient
\begin{equation}\label{eq-MMSE coefficient}
\alpha_{MMSE}=\frac{P\b{h}^T\b{a}}{1+P\|\b{h}\|^2},
\end{equation}
which results in a computation rate region of
\begin{equation}\label{equ-Computation rate with MMSE}
\cR(\b{h},\b{a})= \frac{1}{2} \log^+ \left(\frac{1+P\|\b{h}\|^2}{\|\b{a}\|^2+P\left(\|\b{a}\|^2\|\b{h}\|^2-(\b{h}^T\b{a})^2\right)} \right).
\end{equation}
\end{theorem}
Note that the above theorems are for real channels and the rate expressions for the complex channel are twice the above (\cite[Theorems 3 and 4]{nazer2011compute}). In addition, one should note that the coefficient vector $\b{a}$ must satisfy, 
\begin{equation}\label{equ-Search domain for the vector coefficients}
\|\b{a}\|^2 \leq 1+P\|\b{h}\|^2,
\end{equation}
so that computation rate in \eqref{equ-Computation rate with MMSE} would not be zero ({\cite[Lemma 1]{nazer2011compute}}). 
\begin{remark}[The value of $P$]
We note that the restriction in \eqref{equ-Search domain for the vector coefficients} force a minimal value for the transmission power to employ the CF coding scheme with respect to the coefficient vectors which the decoder chooses.
\end{remark}
\begin{remark}[Computation of several equations]\label{rem-Computation of several equations}
Since the user is free to choose the coefficient vector, $\b{a}$, as he wishes (under the restriction in \eqref{equ-Search domain for the vector coefficients}), he can decode several linear combinations with respect to chosen coefficient vectors $\b{a}_1,\b{a}_2,...$ \emph{from the same transmission} at the expense of reducing the achievable rate. That is, the messages' rates must comply with the lowest computation rate with respect to $\b{a}_1,\b{a}_2,...\ $. Using this technique the user can acquire enough independent linear combinations to retrieve the transmitted message. This technique was shown to achieve the sum capacity of the K-user Gaussian MAC up to a certain gap \cite{ordentlich2014approximate} and was later shown to achieve the entire MAC capacity for the 2-user MAC under specific SNR requirements \cite{zhu2016gaussian}.
\end{remark}

\begin{remark}[AWGN MAC as a special case]
We note that the CF coding scheme can be used also in the AWGN MAC model with no fading, where the messages do not endure any attenuation factors. That is, the messages are aligned together (in a trivial linear combination) at the user and we can use the computation rate region defined in \eqref{equ-Computation rate with MMSE} to determine the achievable rates. 
\end{remark}

\section{PIR for the AWGN MAC}\label{sec-PIR for AWGN MAC}

In this section, we present a retrieval scheme for the AWGN MAC without fading. As mentioned in Section \ref{sec-system model}, we assume that $h_k=1$ for $k=1,...,N$ resulting in the following received signal at the user,
\begin{equation}
\b{y}=\sum_{k=1}^{N}\mathbbm{1}_k\b{x}_k+\mathbf{z},
\end{equation}
where $\mathbbm{1}_k$ equals $1$ if server $k$ is transmitting and $0$ otherwise, as the retrieval scheme may not need all servers. For example, in \cite{banawan2019noisy}, the capacity of the MAC-PIR was achieved by using only 2 servers out of the possible $N$. This is because in the additive modulo-2 MAC, the transmissions of all servers result in a single bit which can then be flipped with probability $q$. Thus, the sum capacity of the additive modulo-2 MAC is equal to the capacity of a Point to Point Binary Symmetric Channel (BSC(q)), which is $1-H(q)$. As a result, the PIR rate in \cite[Theorem 3]{banawan2019noisy} is optimal, and the privacy is attained for "free". However, this is not the case in the AWGN MAC, as will be shown below. In fact, the PIR rate is an increasing function of the number of servers. Furthermore, when we consider fading channels in Section \ref{sec-Joint PIR scheme for the block-fading AWGN - MAC}, we will see that letting all servers transmit is not necessarily optimal.  

We start our analysis by providing a PIR achievability result using separation between the PIR scheme and the channel coding scheme. This achievable PIR rate will constitute a lower bound in later comparison.

\subsection{An Achievable PIR Scheme by Separation}\label{sec-An Achievable PIR scheme by separation}
The PIR scheme presented in \cite{sun2017capacity} requires $N$ noiseless orthogonal channels between the servers and the user. Thus, by using a MAC capacity-achieving code, with which each server encodes his $d$ symbols, one virtually creates such a setting. This is the essence of a separation scheme. Specifically, each server transmits a codeword of length $n$ (channel uses), the user receives the mixed noisy signal and decodes \emph{each server's answer from it}. Thus, the user receives $D=Nd\log{p}$ bits with a sum-rate of $R_{SR}^s= \frac{Nd\log{p}}{n}=\frac{D}{n}$ which can be arbitrarily close to the MAC sum-capacity, $C_{SR}$, as $n$ grows. Accordingly, the PIR rate (in bits per channel use) can be upper bounded by,
\begin{equation*}
R_{PIR}^s=\frac{L\log{p}}{n}=\frac{L\log{p}}{\frac{D}{R_{SR}^s}}=\frac{L\log{p}}{D}R_{SR}<C_{PIR}\cdot C_{SR}.
\end{equation*}
Remembering that $C_{SR}=\frac{1}{2}\log{(1+NP)}$ is the sum-capacity of the AWGN MAC we have,
\begin{equation}\label{equ-achievable PIR rate MAC using separation}
\begin{aligned}
R_{PIR}^s&<\frac{\left(1-\frac{1}{N}\right)}{\left(1-\left(\frac{1}{N}\right)^M\right)}\cdot \frac{1}{2}\log{(1+NP)}\left[\frac{\text{Bits}}{\text{Ch. use}}\right].
\end{aligned}
\end{equation}
We note that due to the additive channel, the rate is measured by the total received bits per channel use at the user, regardless of the fact that each server transmits $n$ symbols individually. This measure is the acceptable metric for such channels since the number of channel uses (bandwidth) is the resource usually being allocated to a system, hence if multiple servers use the same resource it is natural to count them as one. Note, however, that this depends on the fact that each server has its own, non-transferrable, power constraint. This separation scheme is always achievable, and will constitute a lower bound on the retrieval rate that can be obtained under this model. 

We now turn to joint schemes, which outperform the separation-based scheme. We start with $2$ servers, i.e., $N=2$. We then extend the scheme and results to the general case of arbitrary $N$.   

\subsection{A Joint PIR Scheme for Non-Fading AWGN-MAC, $N=2$}

The following theorem presents an achievable retrieval rate for the AWGN MAC. 
\begin{theorem}\label{the- retrieval rate for non-fading MAC}
\textit{For the 2 servers AWGN MAC, the following PIR rate is achievable,
\begin{equation}\label{equ-retrieval rate for the AWGN MAC}
R_{PIR}^J=\frac{1}{2}\log^+{\left(\frac{1}{2}+P\right)}.
\end{equation}} %Thus, the normalized PIR rate is lower bounded by,
%\begin{equation}
%R^* \geq \frac{\log{\left(\frac{1}{2}+P\right)}}{2+\log{\left(\frac{1}{2}+P\right)}}.
%\end{equation}}
%\FMOR{Maybe I should leave it as is with equality as the definition and not bound it...Ask Asaf.}
\end{theorem}

The proof of Theorem \ref{the- retrieval rate for non-fading MAC}, given below, provides a simple and basic scheme for the PIR problem for the 2-servers Gaussian MAC by exploiting the additive nature of the channel. Under this scheme, the servers perform a simple task of computation and the user only performs single-user decoding. This is opposed to the separated solution for this problem (described in Section \ref{sec-An Achievable PIR scheme by separation}, using the result of \cite{sun2017capacity}) where the user needs to send complex structured queries and jointly decode all answers. In addition, we would like to point out that the PIR achievable rate in this scheme does not depend on the number of messages $M$ (a similar observation was made also in \cite{banawan2019noisy} for the MAC-PIR and in \cite{sun2018capacity}).

To gain additional insight on the above result, we compare it to the achievable rate of the separation scheme given in \eqref{equ-achievable PIR rate MAC using separation}. That scheme transforms the MAC into "bit-pipes" such that the user can decode the servers' answers separately. I.e., the user disregards the ability of the channel to compute the sum of the servers' answers, and performs the sum by himself. Thus, any retrieval rate, which is a function of the servers sum-rate, is \emph{constrained by the MAC capacity region}. In addition, the rate is also constrained by the capacity of the PIR scheme, $C_{PIR}$, which is upper bounded in this case by $\frac{2}{3}$ when setting $N=2$ and $M=2$. That is, the separation scheme's rate is strictly lower than the full sum-rate capacity of the MAC. 

\begin{figure}[t]
\centering
    \includegraphics[width=0.6\textwidth]{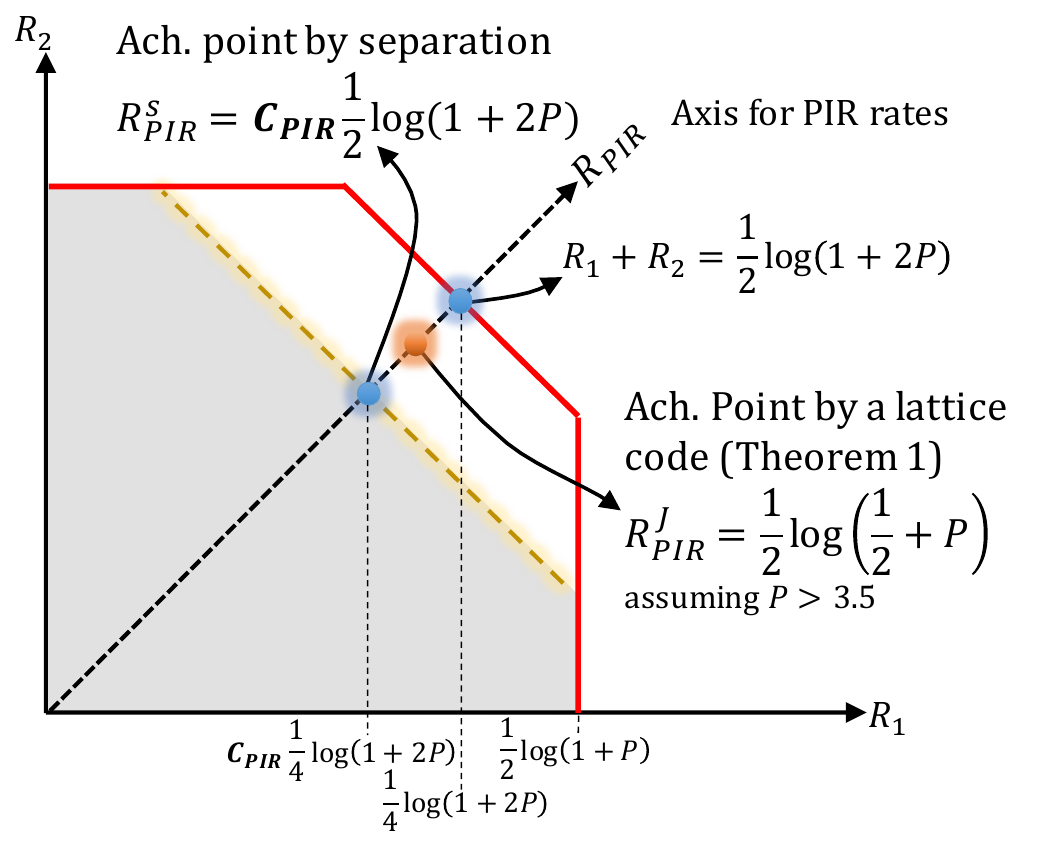}
\caption{The symmetric capacity region for the 2-servers Gaussian MAC for which achievable PIR rates of the private message are shown. On the $R_1=R_2$ line one can see the achievable rate for the PIR scheme suggested in Theorem \ref{the- retrieval rate for non-fading MAC}, which uses lattice codes to decode a sum of two codewords and an achievable rate for a PIR scheme which uses MAC codes, i.e., separates the channel coding and the PIR scheme.}
\label{fig-MAC_region}
\end{figure}

Figure \ref{fig-MAC_region} depicts the PIR rate with respect to the symmetric capacity region of the 2-servers Gaussian MAC (red boundary). That is, any point inside the MAC region describes the rates of the servers for which two messages (without privacy constraint) can be reliably decoded by a receiver. Furthermore, since both servers' answers are in $\mathbb{F}_p^L$ with an equal number of messages, their transmission rates are equal and are located on the symmetric line $R_1=R_2$ inside this region. This line also represents the axis of the PIR rate. That is, this is the actual rate for the private message received at the user. The intersection point (the upper blue dot) between the line $R_1=R_2$ and the capacity region describes the maximal rates the two servers can transmit, resulting in a sum-rate of $\frac{1}{2}\log{(1+2P)}$. However, due to the privacy constraint, the achievable private rate is reduced by $C_{PIR}$, to the lower blue dot. 

% in fact, the \emph{sum of any pair of symmetric rates} located on the line $R_1=R_2$ inside the MAC capacity region \emph{times $C_{PIR}$} (see Figure \ref{fig-MAC_region}). This is due to the fact that both servers' answers are in $\mathbb{F}_p^L$ with an equal number of messages, thus they should have equal rates and since we are restricted to the capacity of the PIR scheme. 
%Accordingly, under this separation scheme the PIR rate is 
%\begin{equation*}
%R_{sep}=C_{PIR}\cdot C_{SR}=\frac{\left(1-\frac{1}{N}\right)}{\left(1-\left(\frac{1}{N}\right)^M\right)}\cdot \frac{1}{2}\log{(1+2P)}
%\end{equation*}
On the other hand, using the retrieval scheme suggested in Theorem \ref{the- retrieval rate for non-fading MAC}, where the user \emph{decodes only a function} of these answers, we are not bound by the MAC region per server, since we do not wish to decode each separately. Hence, the achievable PIR rate, which is still a point on the same symmetric line, can be higher or lower than the achievable point by separation, depending on the value of $P$. This can be shown when comparing \eqref{equ-achievable PIR rate MAC using separation} and \eqref{equ-retrieval rate for the AWGN MAC} as a function of $P$. That is, for low SNR, the separation scheme performs better than the lattice based retrieval scheme. Moreover, considering the restriction in \eqref{equ-Search domain for the vector coefficients}, when $P<1/2$ only the separation scheme (among the two) can achieve a non-zero retrieval rate. Yet, for larger values of $P$, the non-separated scheme outperforms the separated one, and, in fact, the difference can grow larger with $P$ up to achieving the full, non-restricted channel capacity when $P\rightarrow \infty$. Accordingly, we have the following corollary.
\begin{corollary}\label{cor-PIR rate low SNR}
\textit{For the 2 servers AWGN MAC the following PIR rate is achievable,
\begin{equation}
R_{PIR}=\max\left\{ C_{PIR} \cdot \frac{1}{2}\log{(1+2P)},\frac{1}{2}\log^+{\left(\frac{1}{2}+P\right)} \right\}.
\end{equation}
}
\end{corollary} 
The proof follows immediately if the servers and the users are allowed to choose the coding scheme according to the SNR regime. In Figure \ref{fig-MAC_region}, we illustrated the above by the orange dot on the symmetric line where we assume that $P>3.5$. We note that the above Corollary essentially shows that the suggested joint scheme does not perform well in the low SNR regime

\begin{IEEEproof}[\underline{Proof of Theorem \ref{the- retrieval rate for non-fading MAC}}]
The user, which is interested in the message $W_i$, generates a random vector $\b{b}$ of length $M$ such that each entry is either $0$ or $1$ with equal probability. %In addition, the user generates a random variable $s$ which is either $1$ or $-1$ with equal probability. 
Then, the user sends the following vectors as queries to the two servers, 
\begin{equation}\label{equ-queries to the two servers}
Q_1(i)=\b{b}, \ \ \ Q_2(i)=-\left(\b{b}+\b{e}_i(\mathbbm{1}_{\{b_i=0\}}-\mathbbm{1}_{\{b_i=1\}})\right).
\end{equation}
%where $\oplus$ is the XOR operation, i.e., the $i$th entry of $\b{b}$ is flipped. \FMOR{talk with Asaf: I think the $s$ is not needed eventually.} 

From the perspective of the servers, each sees a uniform random vector with an element being zero or non-zero with equal probability. Thus, the privacy of the index $i$ is guaranteed. Specifically, following the privacy requirement in \eqref{equ-queries messages and answers are independent of index} for the $j$th server we have,

\begin{equation*}
\begin{aligned}
I(\theta;Q_j(\theta),\b{x}_j(\theta),W_1^M)&=I(\theta;Q_j(\theta))+I(\theta;\b{x}_j(\theta),W_1^M|Q_j(\theta))\\
&\overset{(a)}{=}I(\theta;Q_j(\theta))\\
&= H(Q_j(\theta))-H(Q_j(\theta)|\theta)\\
&\overset{(b)}{=} M-H(Q_j(\theta)|\theta)\\
&\overset{(c)}{=} M-M=0,
\end{aligned}
\end{equation*}
where $(a)$ follows from $\theta \leftrightarrow Q_j(\theta) \leftrightarrow (\b{x}_j(\theta),W_1^M)$ hence $I(\theta;\b{x}_j(\theta),W_1^M|Q_j(\theta))=0$. $(b)$ follows since for $j=1$, the query $Q_1(\theta)$ is an $i.i.d.$ $(\frac{1}{2},\frac{1}{2})$ random vector $\b{b}$, which has entropy equal to $M$; for $j=2$, the distribution of the query $Q_2(\theta)$ remains the same, since only the $i$th entry of $\b{b}$ is affected, and its value remains independent of the other $M-1$ values, with a distribution which is still $(\frac{1}{2},\frac{1}{2})$. $(c)$ is since for $j=1$ the server observes $\b{b}$ which is independent of $\theta$; for $j=2$, knowing the index $i$ still does not affect the probability of receiving a specific realization of $Q_2(i)$, since all are equiprobable with probability $2^{-M}$.

Upon receiving the queries, the servers perform modulo-p addition between the messages which have non-zero in their corresponding entry in $Q_j(i)$, and form their answers. Specifically,
\begin{equation}\label{equ-servers answer formation}
\begin{aligned}
&A_1=\sum_{m=1}^M Q_{1,m}(i)W_m\mod  p, \\ 
&A_2=\sum_{m=1}^M Q_{2,m}(i)W_m\mod  p.
\end{aligned}
\end{equation}
Note that, $A_1+A_2$ is either $W_i$ or $-W_i$ depending on which server received a non-zero in the $i$th position. Note also that the sign is known to the user.

The servers encode their answers $A_1$ and $A_2$ using the Compute and Forward (CF) coding scheme \cite{nazer2011compute}, which uses nested lattice codebooks. %Furthermore, the code structure and the decoder are similar to \cite{nazer2011compute}.
%The scheme enable receivers to decode "noisy" linear combinations of transmitted messages due to the linear property of the lattice code.
Specifically, we construct a nested lattice codebook as in \cite[Section \RomanNumeralCaps{4}.B]{nazer2011compute}, where $\Lambda_c$ and $\Lambda_f$ are a pair of $n$-dimensional lattices with Voronoi regions $\cV_c$ and $\cV_f$, respectively, such that $\Lambda_c$ is a subset of $\Lambda_f$, i.e., $\Lambda_c \subset \Lambda_f$. The coarse lattice $\Lambda_c$ is used as a shaping region which is scaled to suit the power constraint $P$ and the lattice points from the fine lattice $\Lambda_f$ contained within $\cV_c$ of $\Lambda_c$ are used as the codewords. That is, the nested lattice code is given by, $\cC=\{\Lambda_f \cap \cV_c\}$. In addition, there exist a one-to-one mapping function, $\phi(\cdot)$, between a message $A_j\in\mathbb{F}_p^L$ to the elements of $\cC$ \cite[Lemma 5]{nazer2011compute}.

Accordingly, each server is equipped with a CF encoder, $\cE:\mathbb{F}_p^L \rightarrow \R^n$, that maps length-$L$ messages over the finite field to length-$n$ real-valued codewords, $\b{x}_j=\cE(A_j)$. Specifically, let $\b{v}_j$ be a lattice codeword in $\cC$ such that $\phi(A_j)=\b{v}_j$. Each server is given a dither vector $\b{d}_j$ which is generated independently according to a uniform distribution over the Voronoi region $\cV_c$. The dithers are known to the user. Then, each server transmits
\begin{equation*}
\b{x}_j=[ \b{v}_j-\b{d}_j]  \text{ mod } \Lambda_c. 
\end{equation*}
%Thus, during the encoding process the answer $A_j$ is first mapped into a lattice codeword $\b{v}_j$ in $\cC$ then it is dithered to attain $\b{x}_j$ which is uniformly distributed over $\cV_c$. The servers encode their answers using the same lattice codebook and send them on the MAC channel. 
The received input at the user is thus,
\begin{equation}\label{equ-sum of codewords is a codeword}
\begin{aligned}
\b{y}&=\b{x}_1+\b{x}_2+\b{z}.
\end{aligned}
\end{equation} 
From the noisy sum $\b{x}_1+\b{x}_2+\b{z}$ the user tries to decode the sum of the two lattice codewords $\b{v}=[\b{v}_1+\b{v}_2] \ \text{mod} \  \Lambda_c$  instead of decoding each codeword separately and compute the sum by himself. He does this by computing the following, 
\begin{equation*}
\b{s}=[\alpha \b{y}+\b{d}_1+\b{d}_2]  \text{ mod } \Lambda_c,
\end{equation*}
where $\alpha=\frac{2P}{1+2P}$ is the MMSE coefficient in \eqref{eq-MMSE coefficient}, while using the lattice quantizer to retrieve the estimation $\hat{\b{v}}$, i.e., $Q_{\Lambda_f}(\b{s})=\hat{\b{v}}$. By \cite[Theorem 5]{nazer2011compute} the probability of error $P_r(\hat{\b{v}}\neq \b{v})$ tends to zero exponentially in $n$ as long as the lattice rate $R$, i.e.,the transmission rate of each server, satisfies
\begin{equation}\label{equ-rate for decoding the sum of two codewords no fading} 
R \leq \frac{1}{2} \log^+ \left(\frac{1}{2}+P \right).  
\end{equation}

Consequently, the lattice codeword $\b{v}$, is mapped to either $W_i$ or $-W_i$ since $\phi(A_1)=\b{v}_1$, $\phi(A_2)=\b{v}_2$ and since the linear lattice code preserves the linear operations between the codewords and their corresponding messages \cite[Lemma 6]{nazer2011compute}. Thus, the user can retrieve $W_i$.

%The above rate %was given in \cite[Theorem 5]{nazer2011compute}, 
%is given in Equation \eqref{equ-Computation rate with MMSE} with the channel gains being $(1,1)$ and the coefficient vector $\b{a}$ set to be the all 1 vector. %and $\alpha=\frac{2P}{1+2P}$, where $\alpha$ is a scaling factor used at the decoder. 
%We note that this rate is the rate of the decoded linear combination. Upon successful decoding, 

 %We also note that the specific encoding and decoding schemes are omitted due to lack of space. However, they are presented in \cite{nazer2011compute} along with a thorough description for the construction of the nested lattice codebook.
\end{IEEEproof}

The retrieval rate in \eqref{equ-retrieval rate for the AWGN MAC}, which is essentially the rate at which one can decode the sum of two lattice codewords, appeared also under different contexts in \cite{nam2008capacity, wilson2010joint}. Yet, the CF scheme offers a generalization for the computation of any linear combination. We note that the rate in \eqref{equ-retrieval rate for the AWGN MAC} is the best known achievable rate for such a sum computation 
%and although the resulting equivalent channel in \eqref{equ-sum of codewords is a codeword} is the AWGN channel, i.e., $\b{y}=\b{x}^*+\b{z}$, where $\b{x}^*=\b{x}_1+\b{x}_2$, we cannot achieve a rate of $\frac{1}{2}\log{(1+P)}$ 
(for further reading see \cite{zamir2014lattice}).

\subsection{A Joint PIR Scheme For the Non-Fading AWGN-MAC, Arbitrary $N$}
We now provide an achievable PIR scheme for a general system with $N$ servers. We show that with this scheme, the retrieval rate scales the same as the sum-rate of the non-private Gaussian MAC capacity when $N$ increases. That is, \emph{letting more servers transmit improves the PIR rate}, falling behind the non-private capacity by at most 2 bits per channel use, hence achieving asymptotic optimality with $P$. The scheme uses a coding scheme similar to that in Theorem \ref{the- retrieval rate for non-fading MAC}, where the user tries to decode a linear combination of all transmitted answer. However, instead of computing the sum of all answers, the user intelligently choose his queries in order to aggregate the transmissions from several servers and attain a power gain, thus improving the rate. The scheme and its rate are given in the following corollary and its proof.

\begin{theorem}\label{the- retrieval rate for non-fading MAC N servers}
\textit{For the N servers AWGN MAC, the following PIR rate is achievable,
\begin{equation}\label{eq- retrieval rate for non-fading MAC N servers}
R_{PIR}^J=\frac{1}{2}\log^+{\left(\frac{1}{2}+\floor*{\frac{N}{2}}^2P\right)}.
\end{equation}} 
\end{theorem}
\begin{IEEEproof}
The retrieval scheme for the general system consisting $N$ servers, follows the same steps as the proof of Theorem \ref{the- retrieval rate for non-fading MAC}, yet the queries are designed in a way which creates an equivalent two servers channel between the servers and the user. Specifically, the queries sent by the user are designed so that each pair of servers transmits the sum $\b{x}_1+\b{x}_{2}$. This \emph{implicit} coordination between the servers provides an \emph{increase in the computation rate}, since the user expects to decode a sum of only two codewords, amplified by a constant, instead of a general sum of $N$ codewords. Essentially, this means that the codeword representing the sum will now be received with a significantly \emph{scaled-down noise}. This observation is important, since the rate for decoding a general sum of $N$ codewords, given in \eqref{equ-Computation rate with MMSE}, is significantly lower than the rate achieved with this suggested scheme. The reason lies in the number of self noise penalties (and dithers) \cite{nazer2011compute}, that the CF scheme endures (requires), which reduces to $2$ instead of $N$, as will be shown below. 

Similar to the proof of Theorem \ref{the- retrieval rate for non-fading MAC}, the user generates a random vector $\b{b}$ and sends the following queries to the $l$th and $(l+1)$th servers,
\begin{equation*}
Q_{l}(i)=\b{b}, \ \ Q_{l+1}(i)=-\left(\b{b}+\b{e}_i(\mathbbm{1}_{\{b_i=0\}}-\mathbbm{1}_{\{b_i=1\}})\right),
\end{equation*}  
where $l\in \{1,3,5...,2\floor*{\frac{N}{2}}-1\}$. In case $N$ is odd, the $N$th server is ignored. Then, the servers form their answers as in \eqref{equ-servers answer formation} where each server encodes his answer using a nested lattice code with rate $R$. 

Specifically, let $(\Lambda_f, \Lambda_c)$ be a pair of nested lattices such that the coarse lattice $\Lambda_c$ was is with second moment $P$ to meet the power constraint. The code is known to the user and all the servers. In addition, let $\b{d}_1$ and $\b{d}_2$ be two mutually independent dithers which are uniformly distributed over the Voronoi region $\cV_c$. The dithers are also known in advanced to both the servers and the user. The servers are divided into pairs, where each of the servers in pair $k$, $k\in \{1,3,5...,\floor*{\frac{N}{2}}-1\}$, maps its answer using the mapping function $\phi(\cdot)$ to one of the two lattice codewords, $\b{v}_{k,1}$ or $\b{v}_{k,2}$ respectively. The transmitted signals by each pair of servers are thus given by
\begin{equation*}
\begin{aligned}
\b{x}_{k,1}&= [\b{v}_{k,1}-\b{d}_1] \text{ mod } \Lambda_c \\
\b{x}_{k,2}&= [\b{v}_{k,2}-\b{d}_2] \text{ mod } \Lambda_c,
\end{aligned}
\end{equation*}
Note that since the queries are similar across pairs we have, $\b{v}_{k,1}=\b{v}_{1,1}$ and $\b{v}_{k,2}=\b{v}_{1,2}$ for all $k$. Thus, $[\b{v}_{k,1}+\b{v}_{k,2}]\text{ mod }\Lambda_c=\b{v}$ for all $k$, where $\b{v}$ is the lattice codeword for the private message $W_i$. The received input at the user is thus,
\begin{equation*}
\begin{aligned}
\b{y}&=\sum_{k}(\b{x}_{k,1}+\b{x}_{k,2})+\mathbf{z}.
\end{aligned}
\end{equation*}
In order to decode $\b{v}$, the user computes the following
\begin{equation*}
\b{s}=[\alpha'\b{y}+\b{d}_1+\b{d}_2]\mod \Lambda_c
\end{equation*}   
where $\alpha'=\alpha\frac{1}{\floor*{\frac{N}{2}}}=\frac{2P}{\floor*{\frac{N}{2}}^{-2}+2P}\frac{1}{\floor*{\frac{N}{2}}}$. The above reduces to the Modulo-Lattice Additive Noise (MLAN) channel (\cite{erez2004achieving}) as follows,
\begin{align*}
\b{s}&=[\alpha'\b{y}+\b{d}_1+\b{d}_2]\text{ mod } \Lambda_c\\
&=\Big[\alpha' \sum_{k}(\b{x}_{k,1}+\b{x}_{k,2})+\alpha'\mathbf{z} +\b{d}_1+\b{d}_2\Big]\text{ mod } \Lambda_c\\
&=\Big[\alpha' \sum_{k}\big( [\b{v}_{k,1}-\b{d}_1] \text{ mod } \Lambda_c+\\
&\quad\quad [\b{v}_{k,2}-\b{d}_2] \text{ mod } \Lambda_c\big)+\alpha'\mathbf{z} +\b{d}_1+\b{d}_2\Big]\text{ mod } \Lambda_c\\
&\overset{(a)}{=}\Big[\alpha' \sum_{k}\big( [\b{v}_{k,1}+\b{v}_{k,2}] \text{ mod } \Lambda_c-[\b{d}_1+\b{d}_2] \text{ mod } \Lambda_c\big)\\
&\quad\quad +\alpha'\mathbf{z} +\b{d}_1+\b{d}_2\Big]\text{ mod } \Lambda_c\\
&\overset{(b)}{=}\Big[\alpha \left(\frac{1}{\floor*{\frac{N}{2}}}\sum_{k}\b{v}-\frac{1}{\floor*{\frac{N}{2}}} \sum_{k}[\b{d}_1+\b{d}_2] \text{ mod } \Lambda_c\right)+\\
&\quad\quad \alpha'\mathbf{z} +\b{d}_1+\b{d}_2\Big]\text{ mod } \Lambda_c\\
&=\Big[\alpha \left(\b{v}- [\b{d}_1+\b{d}_2] \text{ mod } \Lambda_c\right)+ \alpha'\mathbf{z} +\b{d}_1+\b{d}_2\Big]\text{ mod } \Lambda_c\\
&=\Big[\b{v}- [\b{d}_1+\b{d}_2] \text{ mod } \Lambda_c+ \alpha'\mathbf{z} +\b{d}_1+\b{d}_2 -\\
& \quad\quad(1-\alpha) \left(\b{v}- [\b{d}_1+\b{d}_2] \text{ mod } \Lambda_c\right)\Big]\text{ mod } \Lambda_c\\
&\overset{(c)}{=}\Big[\b{v}+ \alpha'\mathbf{z}- (1-\alpha) \left(\b{v}- [\b{d}_1+\b{d}_2] \text{ mod } \Lambda_c\right)\Big]\text{ mod } \Lambda_c\\
&\overset{(d)}{=}\Big[\b{v}+ \alpha'\mathbf{z}- (1-\alpha) \left(\b{x}_{1,1}+\b{x}_{1,2}\right)\Big]\text{ mod } \Lambda_c\\
&=\Big[\b{v}+ \b{z}_{eq}\Big]\text{ mod } \Lambda_c,
\end{align*}
where $(a)$ and $(c)$ follow from the distributive property of the $\text{mod } \Lambda_c$ operation. $(b)$ is since the sum of codewords of each pair equals to $\b{v}$. $(d)$ is by replacing the term in the right parenthesis with an equivalent term sent by the first pair of servers. Lastly, we define the equivalent noise term $\b{z}_{eq} \triangleq \alpha'\b{z}- (1-\alpha) \left(\b{x}_{1,1}+\b{x}_{1,2}\right)$. Note that $\b{z}_{eq}$ and $\b{v}$ are independent of each other since $\b{v}_{1,1}$ and $\b{v}_{1,2}$ are independent of $\b{z}$, $\b{x}_{1,1}$ and $\b{x}_{1,2}$. Moreover, $\b{v}$ is a fine lattice point in $\Lambda_f\cap\cV(\Lambda_c)$ which is uniformly distributed in $\cV(\Lambda_c)$ according to the crypto lemma \cite{erez2004achieving}. Accordingly, the second moment of $\b{z}_{eq}$ is given by $\sigma^2_{eq}=\EX[\b{z}_{eq}]=\alpha^2\floor*{\frac{N}{2}}^{-2}+(1-\alpha)^22P$ where we can optimize it on $\alpha$. Specifically, denote by $\sigma_{z'}^2=\floor*{\frac{N}{2}}^{-2}$, we have, $\alpha_{opt}=\frac{2P}{\sigma_{z'}^2+2P}$ and the resulting optimal second moment $\sigma^2_{eq,opt}=\frac{2P\sigma_{z'}^2}{2P+\sigma_{z'}^2}$.
The decoding is successful if $Q_{\Lambda_f}(\b{s})=\b{v}$ which will happen with the probability that the effective noise vector is inside the Voronoi region $\cV(\Lambda_f)$. 

Accordingly, we are left to show the coding rate and the existence of appropriately nested lattices, $(\Lambda_f, \Lambda_c)$, so that $\b{v}$ is decoded correctly with arbitrarily low probability of error. For that manner, we can use \cite[Theorem 1]{wilson2010joint}, which is a modified version of \cite[Theorem 5]{erez2004achieving}, by just setting the channel noise in their result. Specifically, we may write $\b{z}_{eq} \triangleq \alpha\b{z}'- (1-\alpha) \left(\b{x}_{1,1}+\b{x}_{1,2}\right)$, where $\b{z}'$ is a Gaussian noise with variance $\sigma_{z'}^2=\floor*{\frac{N}{2}}^{-2}$ and set it in the rate term in \cite[Theorem 1]{wilson2010joint} which provide us our PIR rate,
\begin{equation*}
R=\frac{1}{2}\log{\left(\frac{1}{2}+\floor*{\frac{N}{2}}^2P\right)}.
\end{equation*}

We note that the extension for the $N$ servers model does not impair the privacy requirement \eqref{equ-queries messages and answers are independent of index}. This is because from the perspective of the $j$th server, it does not matter how many servers are transmitting. Furthermore, for $N=2$ we result with the PIR rate in Theorem \ref{the- retrieval rate for non-fading MAC}.

Consequently, as long as all servers use the same nested lattice code with the above rate, the user can decode the noisy sum $\b{x}_1+\b{x}_2$ with probability of error that tends to zero with $n$ and retrieve $W_i$. 
\end{IEEEproof}

\begin{figure*}[t]
    \centering
    \begin{subfigure}[b]{0.48\textwidth}
        \includegraphics[width=0.98\textwidth]{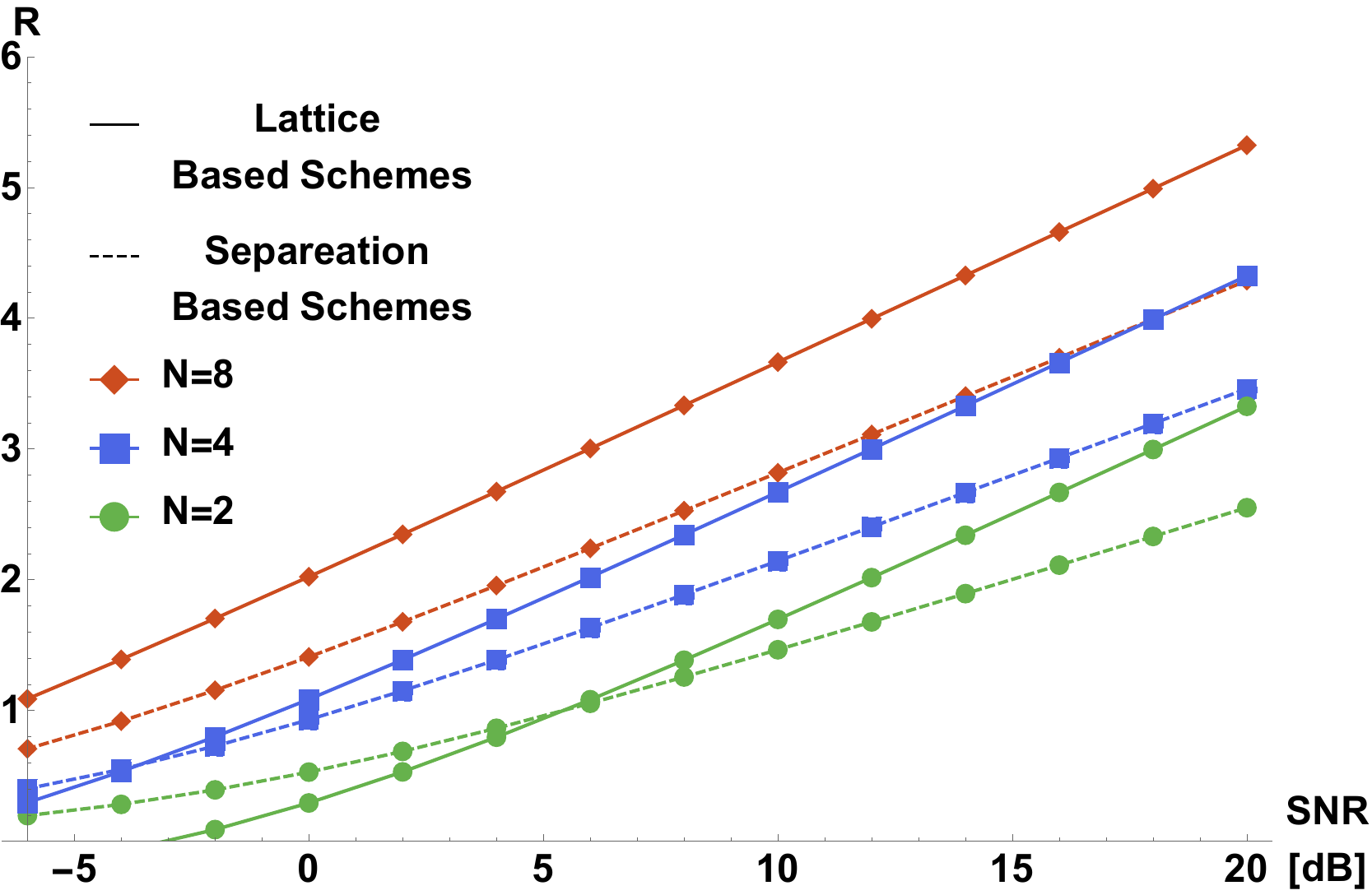}
        \caption{}
        \label{fig-Lower_bound_normalized_rate_no_fading}
    \end{subfigure}
    \begin{subfigure}[b]{0.48\textwidth}
        \includegraphics[width=0.98\textwidth]{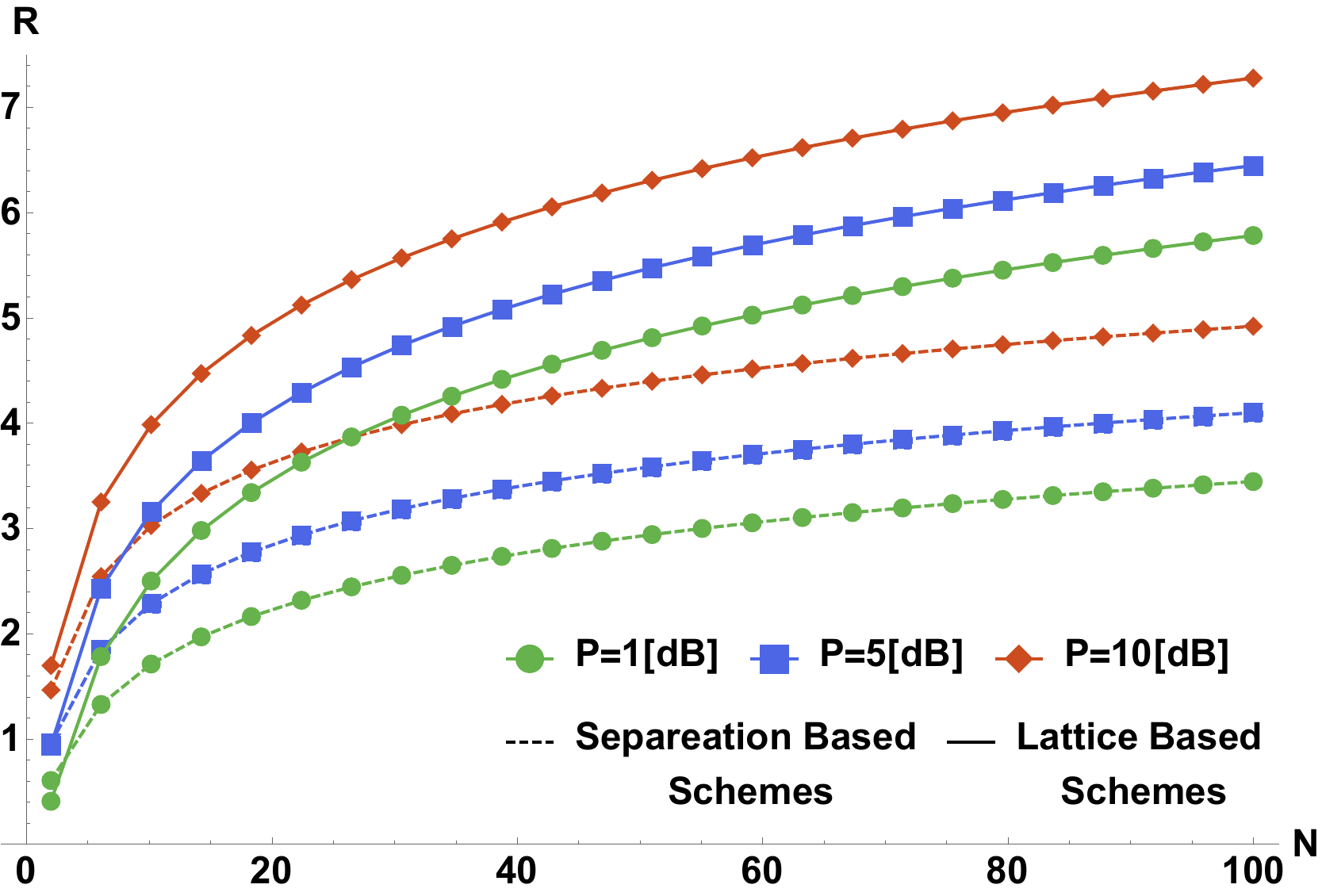}
        \caption{}
        \label{fig-Lower_bound_normalized_rate_no_fading_function_of_N}
    \end{subfigure}
    \caption{The PIR rate as a function of the SNR for $N=\{2,4,8\}$ and as a function of $N$ for $P=\{1\text{dB},5\text{dB},8\text{dB}\}$ are given in $(a)$ and $(b)$ respectively. The dashed lines represent the achievable PIR rate with separation \eqref{equ-achievable PIR rate MAC using separation}. Note that these curves are plotted with $C_{PIR}=\left(1-\frac{1}{N}\right)/\left(1-\left(\frac{1}{N}\right)^M\right)$, given in \cite{sun2017capacity} for 2 messages, i.e., $M=2$, hence \emph{they upper bound the true expressions}. The solid lines are the achievable PIR rates of the lattice based scheme in Theorem \ref{the- retrieval rate for non-fading MAC N servers}.}
    \label{fig-rate as function of N and SNR no fading}
\end{figure*}

Theorem \ref{the- retrieval rate for non-fading MAC N servers} shows that the PIR rate of the joint channel-privacy scheme is an increasing function of $N$ and $P$. Accordingly, one may assess the goodness of this rate when comparing it to the PIR rate of the separation-based scheme and to the full channel capacity without any privacy constraint. While the latter comparison is dealt with in the next subsection, when we explore the gap from the capacity, the following corollary shows that, in the limit of large $N$ and fixed $P$, the suggested joint scheme is twice as good as the separation scheme.
\begin{corollary}\label{cor-corollary twice as good}
When $N\rightarrow\infty$, the ratio between the PIR rate given in Theorem \ref{the- retrieval rate for non-fading MAC N servers} and the PIR rate of the separation-based scheme given in \eqref{equ-achievable PIR rate MAC using separation} is lower bounded by 2. That is,
\begin{equation*}
\lim_{N\rightarrow\infty} \frac{\frac{1}{2}\log^+{\left(\frac{1}{2}+\floor*{\frac{N}{2}}^2P\right)}}{C_{PIR}\frac{1}{2}\log{(1+NP)}} \geq 2.
\end{equation*}
\end{corollary}
\begin{IEEEproof}
We first bound the expression by
\begin{equation*}
 \frac{\frac{1}{2}\log^+{\left(\frac{1}{2}+\floor*{\frac{N}{2}}^2P\right)}}{C_{PIR}\frac{1}{2}\log{(1+NP)}} \geq \frac{\log{\left(\frac{1}{2}+\left(\frac{N-2}{2}\right)^2P\right)}}{\log{(1+NP)}}
\end{equation*}
Taking the limit as $N\rightarrow\infty$, and using L'Hopital's rule, we have
\begin{small}
\begin{equation*}
 \lim_{N\rightarrow\infty} \frac{\log{\left(\frac{1}{2}+\left(\frac{N-2}{2}\right)^2P\right)}}{\log{(1+NP)}} = \lim_{N\rightarrow\infty} \frac{(N-2)(1+NP)}{1+\frac{1}{2}(N-2)^2P} =2.
\end{equation*}
\end{small}
\end{IEEEproof}

Figures \ref{fig-Lower_bound_normalized_rate_no_fading} and \ref{fig-Lower_bound_normalized_rate_no_fading_function_of_N} depict the PIR rate of the lattice based scheme given in Theorem \ref{the- retrieval rate for non-fading MAC N servers} and the PIR rate of the separation scheme given in \eqref{equ-achievable PIR rate MAC using separation}, as a function of the SNR and as a function of the number of servers, respectively. We note that the separation scheme's rate depends on the PIR capacity as given in \cite{sun2017capacity}, which is a function of $M$. For the purpose of comparison we set $M=2$, which upper bounds the capacity expression and thus the separation scheme's rate. 

Specifically, Figure \ref{fig-Lower_bound_normalized_rate_no_fading} shows that even for moderate number of servers the joint scheme is better as the SNR grows. While Figure \ref{fig-Lower_bound_normalized_rate_no_fading_function_of_N} depicts the superiority of the joint scheme when $N$ increase as Lemma \ref{cor-corollary twice as good} suggest. 

From all the above, it is clear that joint privacy and channel encoding and decoding outperforms the separation scheme. Thus, for best performance, for the AWGN MAC, the PIR scheme and the channel coding should be designed together.

%\begin{figure}[t]
%\centering
 %   \includegraphics[width=0.45\textwidth]{PIR_rate_as_function_of_SNR_MAC_no_fading}
%\caption{}
%\label{fig-Lower_bound_normalized_rate_no_fading}
%\end{figure}

%\begin{figure}[t]
%\centering
 %   \includegraphics[width=0.45\textwidth]{PIR_rate_as_function_of_N_MAC_no_fading}
%\caption{The PIR rate as a function of the $N$ for $P=\{1dB,5dB,8dB\}$. The solid lines represent the achievable PIR rate with separation \eqref{equ-achievable PIR rate MAC using separation}. Note that these curves are plotted with $C_{PIR}=\left(1-\frac{1}{N}\right)/\left(1-\left(\frac{1}{N}\right)^M\right)$, given in \cite{sun2017capacity} for 2 messages, i.e., $M=2$, hence \emph{they upper bound the true expressions}. The dashed lines are the achievable PIR rates of the lattice based scheme in Corollary \ref{cor- retrieval rate for non-fading MAC N servers}.}
%\label{fig-Lower_bound_normalized_rate_no_fading_function_of_N}
%\end{figure}

\subsection{Gap From Channel Capacity}\label{sec-Gap from channel capacity}
The results above show that privacy comes with a price of rate reduction, since one must utilize the servers to provide privacy rather than just increasing the transmission rate. That is, privacy does not come for free. However, the price for privacy is not necessarily large, and, in fact, we show that it is bounded. Moreover, this loss can even be negligible in certain scenarios. To quantify this loss, assume that we are not restricted by privacy and consider the capacity in such a model, sending a single message as given in \eqref{equ-MISO sum capacity with per-antenna power constraint}. For the AWGN with no fading, the MISO sum capacity reduces to $\frac{1}{2}\log\left( 1+ N^2P\right)$, which is the maximal rate of transmitting a single message with $N$ servers. Accordingly, the following lemma shows the gap from this capacity.

\begin{lemma}\label{lem-finite gap from channel capacity no fading}
The PIR rate for the $N$ servers AWGN MAC given in Theorem \ref{the- retrieval rate for non-fading MAC N servers} has a finite gap from channel capacity. Namely,
\begin{equation*}
C_{SR}^{MISO}-R_{PIR}^J \leq 2.
\end{equation*}
\end{lemma}
\begin{IEEEproof}
\begin{align*}
 C_{SR}^{MISO}-R_{PIR}^J&=\frac{1}{2}\log\left( 1+ N^2P\right) - \frac{1}{2}\log^+{\left(\frac{1}{2}+\floor*{\frac{N}{2}}^2P\right)}\\
 &\overset{(a)}{\leq}\frac{1}{2}\log\left( \frac{1+ N^2P}{\frac{1}{2}+\floor*{\frac{N}{2}}^2P}\right)\\
 &\leq \frac{1}{2}\log\left( \frac{1+ N^2P}{\frac{1}{2}+\left(\frac{N-1}{2}\right)^2P}\right)\\
 &= \frac{1}{2}\log\left( 4\right)+\frac{1}{2}\log\left( \frac{1+ N^2P}{2+\left( N-1\right)^2P}\right)\\
 &= 1+\frac{1}{2}\log\left( \frac{1+ N^2P}{2+\left( N-1\right)^2P}\right)\\
 &\leq 2,
 \end{align*}
where $(a)$ since $\log^+(x)\geq\log(x)$ and the last inequality follows since $\frac{1+ N^2P}{2+\left( N-1\right)^2P}<4$ for $N\geq2$ for all $P$. 
\end{IEEEproof}
Lemma \ref{lem-finite gap from channel capacity no fading} shows that for every SNR the privacy loss can be upper bounded by no more than 2 bits per channel use. However, we note that in the limit of $P\rightarrow \infty$ and a fixed $N$, the PIR rate is asymptotically optimal as it achieves the capacity.

\section{A Joint PIR Scheme For the Block-Fading AWGN-MAC}\label{sec-Joint PIR scheme for the block-fading AWGN - MAC}
In this section, we present a retrieval scheme for the block-fading AWGN MAC as given in \eqref{equ-the channel}. That is, in contrast to the previous section, the transmitted codewords suffer from attenuation factors and thus do not align trivially to form the integer linear combination $\sum_{k=1}^N\b{x}_k$, as in the non-fading scenario. Yet, the CF coding scheme still enables the user to decode integer linear combinations. This is done by approximating the channel vector $\b{h}$ by an integer coefficient vector $\b{a}$. In what follows, we provide an achievable retrieval scheme for the block-fading AWGN MAC, for a fixed channel vector $\b{h}$, and compute the achievable rate. We then consider Rayleigh fading (i.e., the elements of $\b{h}$ are distributed as a standard normal random variables), and provide 2 achievable schemes with lower bounds on their expected rate. We show that each scheme scales as the expected rate of the capacity with either $P$ or $N$. Alongside this, we present numerical results for the analysis. We compare the results with the average, non-private channel capacity, when the channel's coefficients are known and unknown globally as given in \eqref{equ-MISO sum capacity with per-antenna power constraint} and \eqref{equ-ergodic MISO sum capacity with per-antenna power constraint}, respectively. This is motivated by the understanding that although the channel coefficients are assumed to be unknown by the servers, the user, which sends the queries prior to their transmission, can either include this information (or part of it) in the query or send CSI data implicitly through the queries. That is, the PIR scheme implicitly creates corporation and exploitation of the CSI at the servers. The following is the main result in this section.

\begin{theorem}\label{the-retrieval rate for fading MAC}
\textit{Consider an N servers, block-fading AWGN MAC. Then, for any non-empty subsets of servers $\cS_1,\cS_2$ such that $\cS_1\cap\cS_2=\emptyset$, $\cS_1\cup\cS_2\subseteq\{1,...,N\}$ and some integer vector $\b{a}\in \Z^2$ with non-zero entries, the following PIR rate is achievable,
\begin{equation}\label{equ-retrieval rate for the fading AWGN MAC}
R_{PIR}^J= \frac{1}{2} \log^+ \left(\frac{1+P\|\tilde{\b{h}}\|^2}{\sqn{a}+P\left(a_1\tilde{h}_2-a_2\tilde{h}_1\right)^2} \right),
\end{equation}
where, $\tilde{\b{h}}=(\tilde{h_1},\tilde{h_2})\in \R^2$ and $\tilde{h_1}=\sum_{k\in\cS_1}h_k,\tilde{h_2}=\sum_{k\in\cS_2}h_k$.} 
\end{theorem}
\begin{IEEEproof}
Similar to Theorem \ref{the- retrieval rate for non-fading MAC N servers}, each server encodes his answer using a nested lattice code with rate $R$ to be determined later. However, the queries structure and the assignment to whom they are being sent is different, and depends on the channel vector $\b{h}$.

Specifically, the user divides the servers into 2 non-intersecting subsets, denoted as $\cS_1$ and $\cS_2$. The user sends the query $Q_1(i)$ to each member in $\cS_1$, and $Q_2(i)$ to each member in $\cS_2$. The queries are given in the sequel. Since all servers in subset $\cS_j$ receive the query $Q_j(i)$, they generate the same answer, $A_j$, which is, following the CF scheme, encoded to $\b{x}_j$ (a dithered version of the lattice codeword $\b{v}_j$). Thus, we may write the channel output as follows,
\begin{equation}\label{eq-channel output after subset devision}
\begin{aligned}
\b{y}&=\sum_{k\in\cS_1}h_k\b{x}_1+\sum_{k\in\cS_2}h_k\b{x}_2+\mathbf{z}\\
&=\tilde{h_1}\b{x}_1+\tilde{h_2}\b{x}_2+\mathbf{z},\\
\end{aligned}
\end{equation}
where $\tilde{\b{h}}=(\sum_{k\in\cS_1}h_k,\sum_{k\in\cS_2}h_k)$.

Accordingly, the user is able to decode successfully (i.e., with a probability of error that tends to zero with $n$) a linear combination of the two lattice codewords $\b{v}=[a_1\b{v}_1+a_2\b{v}_2] \ \text{mod} \  \Lambda_c$ from the noisy sum $\tilde{h_1}\b{x}_1+\tilde{h_2}\b{x}_2$, if the lattice rate $R$, i.e., the transmission rate of each server, satisfies %(\cite[Theorem 5]{nazer2011compute})
Equation \eqref{equ-Computation rate with MMSE}. In the context of \eqref{eq-channel output after subset devision}, this results in the following, which can be further simplified as
\begin{equation}\label{equ-achievable rate for decoding the linear combination}
\begin{aligned}
R &= \frac{1}{2} \log^+ \left(\frac{1+P\|\tilde{\b{h}}\|^2}{\|\b{a}\|^2+P\left(\|\b{a}\|^2\|\tilde{\b{h}}\|^2-(\tilde{\b{h}}^T\b{a})^2\right)} \right)\\
&=\frac{1}{2} \log^+ \left(\frac{1+P\|\tilde{\b{h}}\|^2}{\sqn{a}+P\left(\sqn{a}\|\tilde{\b{h}}\|^2-(a_1\tilde{h}_1+a_2\tilde{h}_2)^2\right)} \right)\\
&=\frac{1}{2} \log^+ \left(\frac{1+P\|\tilde{\b{h}}\|^2}{\sqn{a}+P\left(a_1\tilde{h}_2-a_2\tilde{h}_1\right)^2} \right),
\end{aligned}
\end{equation}
where $\b{a}$ is the integer coefficient vector of the linear combination.  
%Accordingly, the rate in \eqref{equ-achievable rate for decoding the linear combination} can be further simplified
%\begin{align*}
%R &\leq \frac{1}{2} \log^+ \left(\frac{1+P\|\tilde{\b{h}}\|^2}{\|\b{a}\|^2+P\left(\|\b{a}\|^2\|\tilde{\b{h}}\|^2-(\tilde{\b{h}}^T\b{a})^2\right)} \right)\\
%&\overset{(a)}{=}\frac{1}{2} \log^+ \left(\frac{1+P\|\tilde{\b{h}}\|^2}{2+P\left(2\|\tilde{\b{h}}\|^2-(a_1\tilde{h}_1+a_2\tilde{h}_2)^2\right)} \right)\\
%&=\frac{1}{2} \log^+ \left(\frac{1+P\|\tilde{\b{h}}\|^2}{2+P\left(a_1\tilde{h}_1-a_2\tilde{h}_2\right)^2} \right)\\
%&\overset{(b)}{=}\frac{1}{2} \log^+ \left(\frac{1+P\|\tilde{\b{h}}\|^2}{2+P\left(|\tilde{h}_1|-|\tilde{h}_2|\right)^2} \right),
%\end{align*}
%where in $(a)$ $\b{a}\in\b{a}^{\b{1}}$ and in $(b)$ we chose $\b{a}$ according to the signs of $\tilde{\b{h}}$. That is, the choice of $\b{a}$ is a deterministic function of the subset devision. We note that any other vector from $\b{a}^{\b{1}}$ can be used also to approximate $\tilde{\b{h}}$, however, this will result with a lower rate\footnote{One can observe in the rate expression that in order to maximize the rate, the signs of the coefficients vector must match (or be opposite to) the channel vector sings. See \cite[Lemma 1]{shmuel2018asymptotically} for more details.}. 

Since the user eventually decodes successfully the linear combination $\b{v}=[a_1\b{v}_1+a_2\b{v}_2] \ \text{mod} \  \Lambda_c$,  the queries must be designed such that $W_i$ can be retrieved from $\b{v}$. Thus, given the subsets $\cS_1$ and $\cS_2$, the user computes the vector $\tilde{\b{h}}$ and chooses the coefficient vector $\b{a}$. Note that the user is not restricted to a certain coefficient vector $\b{a}$, however, a good choice of $\b{a}$, a one that approximates well the channel coefficients will lead to a higher PIR rate. We will engage this issue in the sequel. Moreover, we require that both the entries of $\b{a}$ are non-zero so $W_i$ can be retrieved. Accordingly, the user sets 
\begin{equation}\label{equ-queries to the two servers fading equal sign}
Q_1(i)=q_1^{-1}\b{b}, \ \ Q_2(i)=-q_2^{-1}\left(\b{b}+\b{e}_i(\mathbbm{1}_{\{b_i=0\}}-\mathbbm{1}_{\{b_i=1\}})\right),
\end{equation}
where $q_j=g^{-1}([a_j]\ \text{mod } p)$ are the corresponding coefficients of $\b{a}$ over the prime-sized finite field $\F_p$, i.e., $q_j \in \F_p$, and $g^{-1}(\cdot)$ is a function that maps between the integers $\{0,1,...,p-1\}$ to their corresponding elements in $\F_p$\footnote{We follow the notation of \cite{nazer2011compute} for describing linear combinations.}. Note that the scaling is over the finite field messages and soed affect the transmission power.

The servers form their answers as in \eqref{equ-servers answer formation} which are mapped, using the function $\phi(\cdot)$, to the lattice codewords $\b{v}_1$ and $\b{v}_2$, i.e., $\phi(A_j)=\b{v}_j$. Thus, following \cite[Lemma 6]{nazer2011compute}, the user can retrieve the private message by,
\begin{equation*}
\phi^{-1}(v)=q_1A_1+q_2A_2=\pm W_i,
\end{equation*}
where the sign depends on which query received a non-zero in the $i$th position which is known to the user.

%in case $sign(\tilde{h}_1)=sign(\tilde{h}_2)$. Otherwise, he set,
%\begin{equation}\label{equ-queries to the two servers fading not equal sign}
%Q_1(i)=\b{b}, \ \ \ Q_2(i)=\left(\b{b}+\b{e}_i(\mathbbm{1}_{\{b_i=0\}}-\mathbbm{1}_{\{b_i=1\}})\right).
%\end{equation}

Finally, we note that the above does not impair the privacy requirement \eqref{equ-queries messages and answers are independent of index} even if the channel vector $\b{h}$ is globally known. Specifically we have
\begin{equation*}
\begin{aligned}
I(\theta;\b{h},Q_j(\theta),\b{x}_j(\theta),W_1^M)&=I(\theta;Q_j(\theta),\b{x}_j(\theta),W_1^M|\b{h})+I(\theta;\b{h})\\
&=I(\theta;Q_j(\theta)|\b{h})+I(\theta;\b{x}_j(\theta),W_1^M|Q_j(\theta),\b{h})\\
&=I(\theta;Q_j(\theta)|\b{h})+I(\theta,\b{h};\b{x}_j(\theta),W_1^M|Q_j(\theta))-I(\b{h};\b{x}_j(\theta),W_1^M|Q_j(\theta))\\
&\overset{(a)}{\leq} I(\theta;Q_j(\theta)|\b{h})\\
&\overset{(b)}{=}0.
\end{aligned}
\end{equation*}
$(a)$ follows from the Markov chain $(\theta,\b{h}) \leftrightarrow Q_j(\theta) \leftrightarrow (\b{x}_j(\theta),W_1^M)$ where we note that $\theta$ is independent of $\b{h}$ and thus independent of the values $q_1,q_2$. $(b)$ follows since, in fact, we also have $\theta \leftrightarrow \b{h} \leftrightarrow Q_j(\theta)$ which means that given $\b{h}$, $\theta$ and $Q_j(\theta)$ are independent. That is, given $\b{h}$ the values $q_1,q_2$ can be determined and similarly to the proof of Theorem \ref{the- retrieval rate for non-fading MAC} the distribution of the queries is equiprobable.
\end{IEEEproof}

%We note that the user is not restricted to a certain coefficient vector $\b{a}$, however, a poor choice may lead to a significant rate loss. Moreover, the choice should also take into account the ability to retrieve the private message $W_i$ by only decoding $\b{v}$. 

%In our suggested scheme we consider the set $\b{a}^{\b{1}}=\{(1,1), (-1,1), (1,-1), (-1,-1)\}$ as the possible coefficient vectors. Such a choice ensures that there is no need to scale the servers' answers according to the coefficients. Moreover, since the rate in \eqref{equ-achievable rate for decoding the linear combination} is a decreasing function of the squared norm of $\b{a}$ such a choice most likely will provide higher rates. This property was found useful also in \cite{shmuel2018asymptotically} which provides a more thorough analysis. In addition, in \cite{zhu2016gaussian} a similar observation was found also. 

\begin{figure}[t]
\centering
    \includegraphics[width=0.6\textwidth]{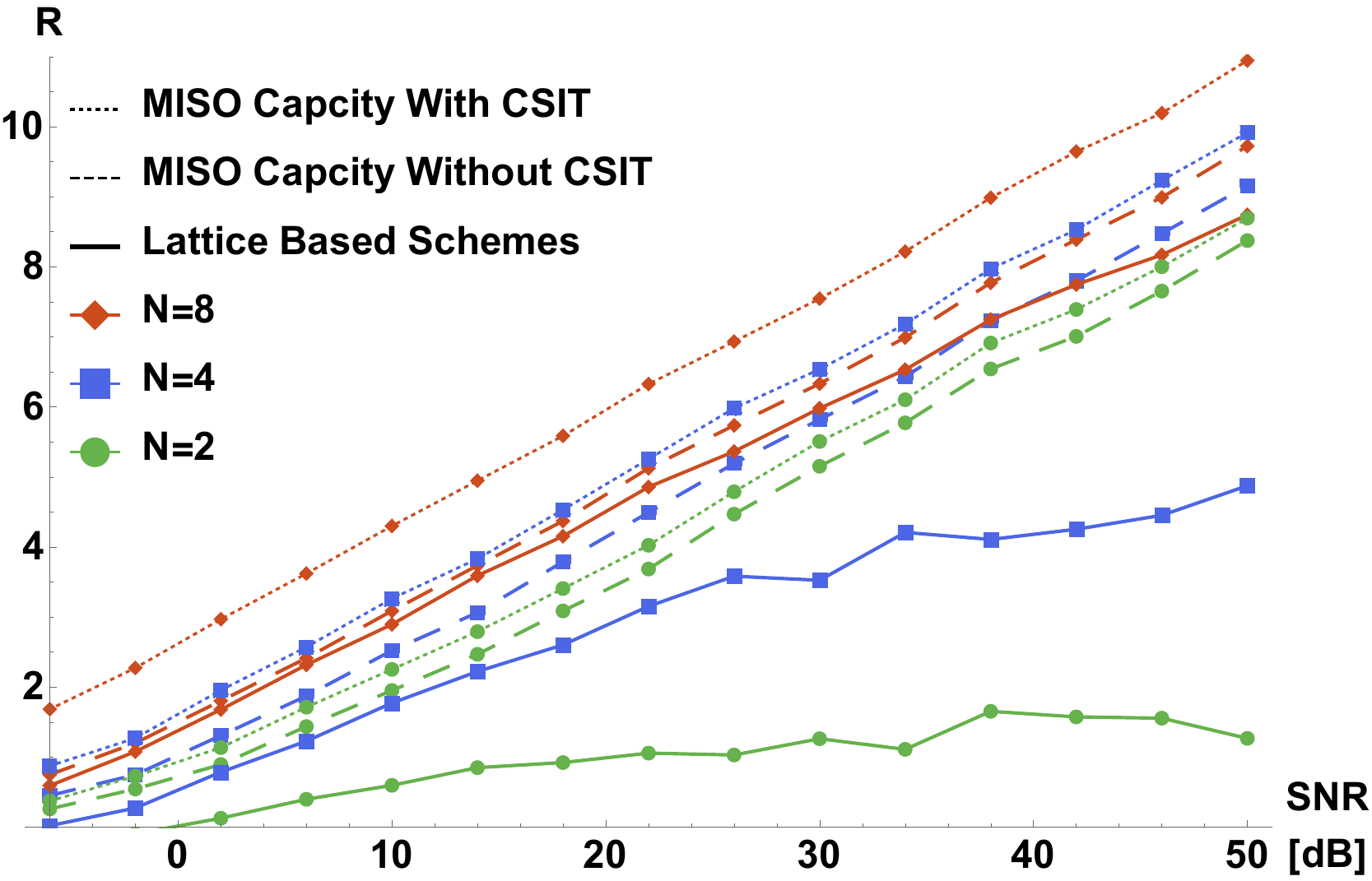}
\caption{The average PIR rate as a function of the SNR for $N=\{2,4,8\}$. The dotted and dashed lines represent the average MISO capacity with and without CSIT as given in \eqref{equ-MISO sum capacity with per-antenna power constraint} and \eqref{equ-ergodic MISO sum capacity with per-antenna power constraint}, respectively. The solid lines are the optimal achievable PIR rate in \eqref{equ-retrieval rate for the fading AWGN MAC maximum} of the lattice based scheme given in Theorem \ref{the-retrieval rate for fading MAC} (a slight weaken version to improve simulation running time).} 
\label{fig-PIR_rate_fading_MISO_capaciy}
\end{figure}

The achievable PIR scheme given in Theorem \ref{the-retrieval rate for fading MAC} is based on the restriction which the user should decode a certain linear combination of only 2 codewords (answers). That is, a devision to only 2 subsets of servers. However, although sub-optimal, this strategy is justified for the following reasons. First, it is inspired from the previous section for non-fading channels, where this strategy is asymptotically optimal. Second, letting the number of subsets grow, the number of different queries sent will grow as well, resulting in a larger linear combination of codewords. Hence, assuming the coding scheme is CF, the computation rate (transmission rate) decreases due to the increasing penalties of the equivalent noise added with every integer coefficient \cite{shmuel2020compute}. Lastly, as numerical results in Figure \ref{fig-PIR_rate_fading_MISO_capaciy} show, this scheme's rate is approaching the capacity as the number of servers grows for every $P$. 

In order to analyze the PIR rate in \eqref{equ-retrieval rate for the fading AWGN MAC}, one should note that the user may choose $\cS_1$, $\cS_2$ and the coefficient vector $\b{a}$ to maximize it. Namely, we have the following global optimization problem,
\begin{equation}\label{equ-retrieval rate for the fading AWGN MAC maximum}
\max_{\substack{\cS_1,\cS_2,\b{a}\\ a_j\neq0}}\left\{\frac{1}{2}\log^+{\left(\frac{1+P\left(\left(\sum\limits_{k\in\cS_1}h_k\right)^2+\left(\sum\limits_{k\in\cS_2}h_k\right)^2\right)}{\sqn{a}+P\left(a_1\sum\limits_{k\in\cS_2}h_k-a_2\sum\limits_{k\in\cS_1}h_k\right)^2}\right)}\right\}.
\end{equation}
Finding the optimal partition and the optimal coefficient vector $\b{a}$, i.e., finding the optimal solution for the above optimization, is a hard problem. Even for a fixed $\b{a}$, the problem relates to the subset sum problem (or partition problem) which is NP-complete \cite{cormen2009introduction}. To maximize the rate, the absolute values of the two sums, multiplied by the corresponding elements of $\b{a}$, should be as close to each other as possible (to minimize the denominator), while being as large as possible. Moreover, the union of the sets is not restricted to contain all servers, and a possible optimal solution can prevent a server from transmitting. However, in the next subsection, we suggest two sub-optimal schemes, with lower bounds on their expected rate. Specifically, each scheme attempts to maximize the rate in \eqref{equ-retrieval rate for the fading AWGN MAC maximum} differently, which results in optimal scaling laws for the regimes, of large $P$ or $N$. In addition, we provide numerical results which are based on a heuristic algorithm to solve the optimization.

\begin{remark}[Comparison with a separation based scheme]
A comparison with a separation based scheme should consider the fact that by using MAC capacity-achieving codes, the "virtual" orthogonal channels between the servers and the user are not symmetric due to the fading. For such a scenario, the exact description of the $C_{PIR}$ is not known in general and only upper and lower bounds are known \cite{banawan2019noisy}. 
\end{remark}

For small values of $N$, the optimization in \eqref{equ-retrieval rate for the fading AWGN MAC maximum} can be solved by exhaustive search. Figure \ref{fig-PIR_rate_fading_MISO_capaciy} depicts simulation results for the average PIR rate as a function of the SNR for different $N$.The solid lines, which are the optimal solutions for the PIR rate, are compared with the average non-private channel capacity with (dotted lines) and without (dashed lines) CSIT. One may observe that the optimal PIR rate is an increasing function of the SNR and as the number of servers grows, the PIR rate approaches the two capacities of the channel for every SNR. That is, the figure shows that the optimal solution for \eqref{equ-retrieval rate for the fading AWGN MAC maximum} scales with $P$ and $N$ simultaneously. Moreover, once the user has more variables for the optimization (i.e., as $N$ grows) significant improvement may be reached. For example, for $N=2$, where such optimization is not possible, the performance is far from the capacity.

\subsection{Lower Bounds on the Expected Achievable Rate}
In this subsection, we provide schemes and their resulting lower bounds on the expected achievable rate, assuming Rayleigh fading. These lower bounds are based on specific sub-optimal choices (sub-optimal solutions for the maximization in \eqref{equ-retrieval rate for the fading AWGN MAC maximum}) for $\cS_1,\cS_2$ and $\b{a}$, i.e., the partition of servers into $2$ non-intersecting subsets, while using the retrieval scheme given in Theorem \ref{the-retrieval rate for fading MAC}. The first bound and the scheme for the partition of servers are given in the following theorem.

\begin{theorem}\label{the-expected rate lower bound}
\textit{The expected PIR rate in \eqref{equ-retrieval rate for the fading AWGN MAC maximum} is lower bounded by,
\begin{equation}\label{equ-expected rate lower bound}
\begin{aligned}
 \EX\left[R_{PIR}^{J,max}\right]&=\EX\left[\max_{\substack{\cS_1,\cS_2,\b{a}\\ a_j\neq0}}\left\{\frac{1}{2}\log^+{\left(\frac{1+P\sqn{\tilde{h}}}{\sqn{a}+P(a_1\tilde{h_2}-a_2\tilde{h_1})^2}\right)}\right\}\right]\\
 &\geq \frac{1}{2}\log{\left(\frac{(4+N'^2Pc)}{4\left(2+N'P\right)}\right)} -o(1),
 \end{aligned}
\end{equation}
where $c=\left(\sqrt{\frac{2}{\pi}}-\frac{1}{2}\right)^2$ and $N'=2\min\{N_1,N_2\}$ where $N_1$ and $N_2$ are the number of positive and negative elements in $\b{h}$, respectively.}
\end{theorem}

First, to be able to assess the scaling law of this lower bounds, the following lemma shows that $N'$ converges to $N$ as $N$ grows.
\begin{lemma}\label{lem-N' converge to N}
Let $\b{h}$ be a random standard Gaussian vector of length $N$. Let $N'=2\min\{N_1,N_2\}$, where $N_1,N_2$ are the number of positive and negative elements, respectively. Then, $N'$ converge to $N$ in $L^2$-norm. That is,
\begin{equation}
N' \overset{L^2}{\rightarrow} N.
\end{equation}
\end{lemma}
\begin{IEEEproof}
\begin{equation*}
\begin{aligned}
N'&=2\min\{N_1,N_2\}\\
&=N_1+N_2-|N_1-N_2|\\
&\overset{(a)}{=}N-|2N_1-N|\\
&\overset{(b)}{=}N\left(1-2\left|\frac{1}{N} \sum_{i=1}^N \mathbbm{1}_{\{h_i>0\}}-\frac{1}{2}\right|\right)
\end{aligned}
\end{equation*}
In $(a)$ $N=N_1+N_2$. $(b)$ $N_1$ is a sum of Bernoulli random variables with probability 0.5 which is the probability of $h_i$ to be positive. Since, $\sum_{i=1}^N \mathbbm{1}_{\{h_i>0\}} \overset{L^2}{\rightarrow} \frac{1}{2}$ the result follows.
\end{IEEEproof}

\begin{IEEEproof}[Proof of Theorem \ref{the-expected rate lower bound}]
Pick $\cS_1$,$\cS_2$ and $\b{a}$ as follows. Let $N_1$ be the number of positive element in $\b{h}$ and let $N_2$ be the number of negative elements in $\b{h}$. Clearly, $N_1+N_2=N$. In addition, let $N'=2\min\{N_1,N_2\}$. We construct $\cS_1$ to be a set of size $\frac{N'}{2}$, chosen uniformly from all positive elements of $\b{h}$, and $\cS_2$ to be a set of size $\frac{N'}{2}$, chosen uniformly from all negative elements of $\b{h}$. Note that either $\cS_1$ or $\cS_2$ will hold \emph{all} the positive or negative elements from $\b{h}$. In the extreme scenario of all elements of $\b{h}$ having the same sign, we choose $\cS_1$ and $\cS_2$ to each have $N'=2\floor*{N/2}$ uniformly chosen elements of $\b{h}$ with no repetition. Given $\cS_1$ and $\cS_2$, compute the vector $\tilde{h}=(\tilde{h_1},\tilde{h_2})$ and set $\b{a}=(sign(\tilde{h_1})\cdot 1,sign(\tilde{h_2}) \cdot 1)$. 

Recall that $\b{a}$ determines the specific linear combination the user decodes. Since the queries in \eqref{equ-queries to the two servers fading equal sign} are constructed such that the linear combination is normalized by the corresponding finite field elements, the choice of $\b{a}$ does not affect on the decoding itself. However, as can be seen in \eqref{equ-retrieval rate for the fading AWGN MAC}, the rate is a decreasing function of the squared norm of $\b{a}$. Thus, setting $\b{a}$ such that its norm has the smallest value may be considered as a good option for the maximization of the rate. We note that, under different but similar context, such a choice for the $\b{a}$ vector was found also as a good solution in \cite{zhu2016gaussian,shmuel2020compute}. 
Accordingly, given the above choice of $\b{a}$, and the fact that the negative and positive elements in absolute value of a Gaussian random vector have the same distribution we choose $\cS_1$ and $\cS_2$ so that on average (and when $N$ is large), $\tilde{h_1}$ and $\tilde{h_2}$ would have similar high value.

%Note that the restriction on the cardinality of the sets $\cS_1$ and $\cS_2$ to be equal to $N'/2$, is for simplicity purposes and the results will behave similarly if we remove it. This is due to Lemma \ref{lem-N' converge to N} which implies that the cardinality of the sets is roughly the same. 
 
We start with the maximization problem in \eqref{equ-retrieval rate for the fading AWGN MAC maximum} where, for ease of notation, we use the vector representation of $\tilde{\b{h}}=(\tilde{h_1},\tilde{h_2})$.
\begin{align}
 &\EX\left[\max_{\substack{\cS_1,\cS_2,\b{a}\\ a_j\neq0}}\left\{\frac{1}{2}\log^+{\left(\frac{1+P\sqn{\tilde{h}}}{\sqn{a}+P(a_1\tilde{h_2}-a_2\tilde{h_1})^2}\right)}\right\}\right]\nonumber\\
 &\overset{(a)}{\geq} \EX\left[\frac{1}{2}\log{\left(\frac{1+P(\tilde{h_1^*}^2+\tilde{h_2^*}^2)}{2+P(|\tilde{h_1^*}|-|\tilde{h_2^*}|)^2}\right)}\right] \nonumber\\
 &= \EX\left[\frac{1}{2}\log{\left(1+P(\tilde{h_1^*}^2+\tilde{h_2^*}^2)\right)}\right]\nonumber-\EX\left[\frac{1}{2}\log{\left(2+P\left(|\tilde{h_1^*}|-|\tilde{h_2^*}|\right)^2\right)}\right]\nonumber\\
 &\overset{(b)}{\geq} \EX\left[\frac{1}{2}\log{\left(1+P(\tilde{h_1^*}^2+\tilde{h_2^*}^2)\right)}\right]
 %&\quad\quad\quad\quad\quad\quad-\frac{1}{2}\log{\left(2+P\EX\left[\left(\tilde{h_1^*}+\tilde{h_2^*}\right)^2\right]\right)}\\
 %&= \EX\left[\frac{1}{2}\log{\left(1+P(\tilde{h_1^*}^2+\tilde{h_2^*}^2)\right)}\right]\\
-\frac{1}{2}\log{\left(2+P\EX\left[\left(\sum_{k\in\cS_1^*\cup\cS_2^*}h_k\right)^2\right]\right)}\label{eq-factor to P}\\
&\overset{(c)}{=}\EX\left[\frac{1}{2}\log{\left(1+(\tilde{h_1^*}^2+\tilde{h_2^*}^2)P\right)}\right]-\frac{1}{2}\log{\left(2+N'P\right)}\nonumber\\
&\overset{(d)}{\geq}\EX\left[\frac{1}{2}\log{\left(1+\tilde{h_1^*}^2P\right)}\right]-\frac{1}{2}\log{\left(2+N'P\right)}.\label{eq-first break}
\end{align}
$(a)$ follows from the suboptimal choice for $\cS_1$, $\cS_2$ and $\b{a}$ as given above where we denote this choice by $(\cdot)^*$. In addition, note that $\log^+(x)\geq\log(x)$. $(b)$ follows from Jensen's inequality and the fact that the sign of $\tilde{h_1^*}$ is positive and the sign of $\tilde{h_2^*}$ is negative; thus, $|\tilde{h_1^*}|-|\tilde{h_2^*}|=\tilde{h_1^*}+\tilde{h_2^*}$. $(c)$ follows since the sum in the second term is on $N'$ $i.i.d.$ standard Gaussian random variables. %In $(d)$, since $\tilde{h_1^*}^2,\tilde{h_2^*}^2$ are $i.i.d.$ random variables, we can bound their sum by twice the lowest value among them. Without loss of generality, we assume that $\tilde{h_1^*}^2\leq\tilde{h_2^*}^2$.
We now lower bound the first term as follows,

\begin{align*}
\EX\left[\frac{1}{2}\log{\left(1+\tilde{h_1^*}^2P\right)}\right]&= \EX\left[\frac{1}{2}\log{\left(1+\tilde{h_1^*}^2P\right)}\Big| \left|\frac{2}{N'}\tilde{h_1^*}-\sqrt{\frac{2}{\pi}}\right|\leq\epsilon \right]\\
&\quad \quad \quad \quad\quad \quad \quad\quad \quad \quad \quad \quad \cdot P_r\left(\left|\frac{2}{N'}\tilde{h_1^*}-\sqrt{\frac{2}{\pi}}\right|\leq\epsilon\right)\\
&\quad + \EX\left[\frac{1}{2}\log{\left(1+\tilde{h_1^*}^2P\right)}\Big| \left|\frac{2}{N'}\tilde{h_1^*}-\sqrt{\frac{2}{\pi}}\right|>\epsilon \right]\\
&\quad \quad \quad \quad\quad \quad \quad\quad \quad \quad \quad \quad \cdot P_r\left(\left|\frac{2}{N'}\tilde{h_1^*}-\sqrt{\frac{2}{\pi}}\right|>\epsilon\right)\\
&\geq  \EX\left[\frac{1}{2}\log{\left(1+\tilde{h_1^*}^2P\right)}\Big| \left|\frac{2}{N'}\tilde{h_1^*}-\sqrt{\frac{2}{\pi}}\right|\leq\epsilon \right]\\
&\quad \quad \quad \quad\quad \quad \quad\quad \quad \quad \quad \quad \cdot P_r\left(\left|\frac{2}{N'}\tilde{h_1^*}^2-\sqrt{\frac{2}{\pi}}\right|\leq\epsilon\right)\\
&\overset{(e)}{\geq}  \frac{1}{2}\log{\left(1+\frac{N'^2P}{4}\left(\sqrt{\frac{2}{\pi}}-\epsilon\right)^2\right)} P_r\left(\left|\frac{2}{N'}\tilde{h_1^*}-\sqrt{\frac{2}{\pi}}\right|\leq\epsilon\right)\\
&\overset{(f)}{\geq}  \frac{1}{2}\log{\left(1+\frac{N'^2P}{4}\left(\sqrt{\frac{2}{\pi}}-\epsilon\right)^2\right)} \left(1-\frac{\text{Var}\left(\frac{2}{N'}\tilde{h_1^*} \right)}{\epsilon^2}\right)\\
&\overset{(g)}{=}  \frac{1}{2}\log{\left(1+\frac{N'^2P}{4}\left(\sqrt{\frac{2}{\pi}}-\epsilon\right)^2\right)} \left(1-\frac{2\left(1-\frac{2}{\pi} \right)}{N'\epsilon^2}\right)\\
&=  \frac{1}{2}\log{\left(1+\frac{N'^2Pc}{4}\right)} -o(1).\\
\end{align*}
$(e)$ follows since $\left|\frac{2}{N'}\tilde{h_1^*}-\sqrt{\frac{2}{\pi}}\right| \leq \epsilon$ and thus $\tilde{h_1^*}\geq \frac{N'}{2}\left(\sqrt{\frac{2}{\pi}}-\epsilon\right)$. $(f)$ follows from Chebyshev's inequality where we require that $\sqrt{\frac{2}{N'}\left(1-\frac{2}{\pi}\right)}<\epsilon<\sqrt{\frac{2}{\pi}}$. Any $\epsilon$ outside this interval will lead to a meaningless result. Thus, we set it to be $\epsilon=0.5$ and we denote $c=\left(\sqrt{\frac{2}{\pi}}-\frac{1}{2}\right)^2$. 
In $(g)$, $Var(\tilde{h_1^*})=Var(\sum_{k\in\cS_1}h_k)=\frac{N'}{2}\left(1-\frac{2}{\pi} \right)$ since the elements in $\cS_1$ are $i.i.d.$ random variables distributed as Half-Normal distribution with mean $\sqrt{\frac{2}{\pi}}$ and variance $1-\frac{2}{\pi}$.

Accordingly, setting the above derivation in \eqref{eq-first break}, we have
\begin{align*}
&\geq \frac{1}{2}\log{\left(1+\frac{N'^2Pc}{4}\right)}-\frac{1}{2}\log{\left(2+N'P\right)}-o(1)\\ 
&=\frac{1}{2}\log{\left(\frac{(4+N'^2Pc)}{4\left(2+N'P\right)}\right)} -o(1).
\end{align*}
\end{IEEEproof}

\begin{figure}[t]
\centering
    \includegraphics[width=0.6\textwidth]{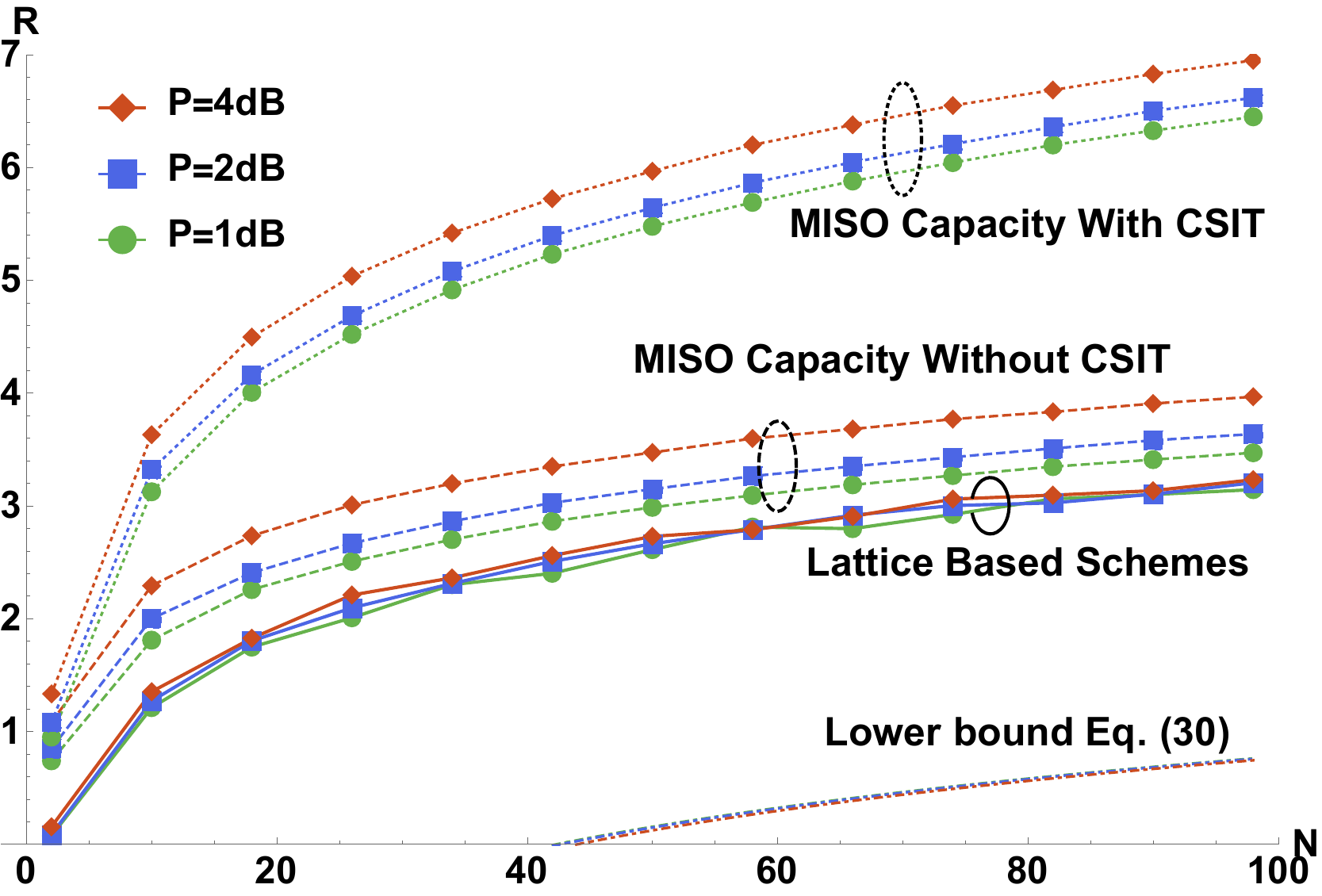}
\caption{Simulation results of the average PIR rate (solid lines) of the scheme given in Theorem \ref{the-expected rate lower bound} with the analytical lower bound (dashed-doted line) in \eqref{equ-expected rate lower bound} as a function of $N$ for $P=\{1\text{dB},2\text{dB},4\text{dB}\}$. The dotted and dashed lines represent the MISO capacity as given in \eqref{equ-MISO sum capacity with per-antenna power constraint} and \eqref{equ-ergodic MISO sum capacity with per-antenna power constraint} for the cases of known and unknown CSIT, respectively.}
\label{fig-rate as function of N}
\end{figure}

Theorem \ref{the-expected rate lower bound} suggests that the PIR rate is an increasing function of the number of servers, and, in fact, it scales like $O(\log(N))$ which is the scaling law of the capacity without CSIT. Figure \ref{fig-rate as function of N} depicts simulation results (solid line) for the PIR scheme suggested in Theorem \ref{the-expected rate lower bound} and the lower bound (dashed-doted line) in \eqref{equ-expected rate lower bound} as a function of $N$. These results are compared with the MISO capacity as given in \eqref{equ-MISO sum capacity with per-antenna power constraint} and \eqref{equ-ergodic MISO sum capacity with per-antenna power constraint} for the cases of known and unknown CSIT, respectively. Note that the MISO capacity with CSI is a loose upper bound in any case, since not only CSI may be only implicitly received through the queries, it is bounded by $M$ bits, the size of the query. As mentioned, Figure \ref{fig-rate as function of N} depicts similar scaling law to the MISO capacity, up to a constant gap. However, in this scheme, the increase in $P$ does not provide an increase in the rate. That is, the rate does not scale well with $P$. This can be explained if one recalls the expression in \eqref{equ-retrieval rate for the fading AWGN MAC maximum}. The choice of positive and negative groups will, on one hand, maximize the expectation of the numerator while on the other hand, will not minimize the expectation of the squared difference between the sums. That is, there is a factor of $N'$ to $P$ in the denominator (Equation \eqref{eq-factor to P}) instead of canceling it as $N$ grows. That is, when $P\rightarrow\infty$ the rate becomes a constant. In the following theorem we suggest a different PIR scheme, based on a different partition, which picks two servers which simultaneously minimizes this difference and maximizes the numerator. The result will be optimal scaling with $P$ for fixed $N$.

\begin{theorem}\label{the-expected rate lower bound with P}
\textit{The expected PIR rate in \eqref{equ-retrieval rate for the fading AWGN MAC maximum} is asymptotically lower bounded by,
\begin{equation}\label{equ-expected rate lower bound with P}
\begin{aligned}
 \EX\left[R_{PIR}^{J,max}\right]&=\EX\left[\max_{\substack{\cS_1,\cS_2,\b{a}\\ a_j\neq0}}\left\{\frac{1}{2}\log^+{\left(\frac{1+P\sqn{\tilde{h}}}{\sqn{a}+P(a_1\tilde{h_2}-a_2\tilde{h_1})^2}\right)}\right\}\right]\\
 &\geq  \frac{1}{2}\log{\left(1+2u^2P\right)}-\frac{1}{2} -o\left(\frac{1}{\log{N}}\right),
 \end{aligned}
\end{equation}
where $u= \sqrt{2\ln\frac{2\sqrt{N}}{\ln{N}\sqrt{2 \pi}}}-\frac{1}{\ln{N}}$, that is, $u=\omega(1)$.
}
\end{theorem}
Note that $u$ was chosen to provide an asymptotic lower bound. The suggested PIR scheme, however, is independent of $u$ and can be applied to any number of servers.
\begin{IEEEproof}
Define an interval $\Delta=[-(u+\delta),-u]\cup[u,u+\delta]$ where the values $u(N)$ and $\delta(N)$ will be given in the sequel. In addition, Let $\xi$ denote the event of having at least $2$ elements in $\b{h}$ with values in $\Delta$. Note that $P_r(\xi)$ follows a binomial distribution with probability of success $p(u,\delta)=2(\Phi(u+\delta)-\Phi(u))$ where $\Phi$ is the CDF of the normal distribution. Accordingly, we can lower bound it using the Chernoff's bound, which requires that $Np(u,\delta) \geq 1$, as follows, 
\begin{equation}\label{equ-lower bound on probability xi}
\begin{aligned}
P_r(\xi)&=1- \left((1-p(u,\delta))^N+Np(u,\delta)(1-p(u,\delta))^{N-1}\right)\\
	&\geq 1-e^{-\frac{1}{2p(u,\delta)}\frac{(Np(u,\delta)-1)^2}{N}}.
\end{aligned}
\end{equation}

As already proved usefull for CF \cite{shmuel2020compute}, the PIR scheme chooses two arbitrary servers for transmission such that their channel coefficients are in $\Delta$. That is, the sets $\cS_1$,$\cS_2$ are the two servers picked, one in each set. The probability of finding such two servers is $P_r(\xi)$ which, as will be shown below, tends to one as $N$ grows. In addition, pick $\b{a}=(sign(\tilde{h_1})\cdot 1,sign(\tilde{h_2}) \cdot 1)$ as in Theorem \ref{the-expected rate lower bound}. Accordingly,
\begin{align*}
 \EX\left[\max_{\substack{\cS_1,\cS_2,\b{a}\\ a_j\neq0}}\left\{\frac{1}{2}\log^+{\left(\frac{1+P\sqn{\tilde{h}}}{\sqn{a}+P(a_1\tilde{h_2}-a_2\tilde{h_1})^2}\right)}\right\}\right]
 &\overset{(a)}{\geq} \EX\left[\frac{1}{2}\log{\left(\frac{1+P(\tilde{h_1^*}^2+\tilde{h_2^*}^2)}{2+P(|\tilde{h_1^*}|-|\tilde{h_2^*}|)^2}\right)}\right] \nonumber\\
 & \geq \EX\left[\frac{1}{2}\log{\left(\frac{1+P(\tilde{h_1^*}^2+\tilde{h_2^*}^2)}{2+P(|\tilde{h_1^*}|-|\tilde{h_2^*}|)^2}\right)}\Bigg| \xi \right]P_r(\xi)\\
  &\overset{(b)}{\geq} \frac{1}{2}\log{\left(\frac{1+2u^2P}{2+\delta^2P}\right)}P_r(\xi),
  \end{align*}
where $(a)$ follows from the suboptimal choice for $\cS_1$, $\cS_2$ and $\b{a}$ as given above. In addition, note that $\log^+(x)\geq\log(x)$. $(b)$ follows since $\tilde{h_1^*},\tilde{h_2^*}\in\Delta$. By setting 
\begin{equation}\label{equ-u(L) and delta(L) values}
u(N) =\sqrt{2\ln\frac{2\delta\sqrt{N}}{\sqrt{2 \pi}}}-\delta \quad \text{and} \quad  \delta(N)=\frac{1}{\ln{N}}
\end{equation}
we can lower bound $p(u,\delta)$ as follows,
\begin{equation}\label{equ-p(u,delta) lower bound}
\begin{aligned}
p(u,\delta)&= 2(\Phi(u+\delta)-\Phi(u)) \\
&= \frac{2}{\sqrt{2\pi}} \int_{u}^{u+\delta} e^{-\frac{t^2}{2}}dt\\
&\geq \delta  \frac{2}{\sqrt{2\pi}} e^{-\frac{(u+\delta)^2}{2}}\\
&= \delta  \frac{2}{\sqrt{2\pi}} e^{-\frac{\left(\sqrt{2\ln\frac{\delta\sqrt{N}}{\sqrt{2 \pi}}}-\delta+\delta\right)^2}{2}}\\
&= \frac{1}{\sqrt{N}},
\end{aligned}
\end{equation}
such that the Chernoff's bound requirement is satisfied, i.e., $N p(u,\delta)\geq1$ and the probability $P_r(\xi)$ tends to one with $N$. This can be seen as follows,
\begin{equation}\label{equ-limit of P_r(xi) goes to one}
\begin{aligned}
&\lim_{N \rightarrow \infty} P_r(\xi) \\
& \geq \lim_{N \rightarrow \infty} 1-e^{-\frac{1}{2p(u,\delta)}\frac{(Np(u,\delta)-1)^2}{N}} \\
& \overset{(a)}{\geq} \lim_{N \rightarrow \infty} 1-e^{-\frac{1}{2}\frac{(\sqrt{N}-1)^2}{\sqrt{N}}} \\
&=1,
\end{aligned}
\end{equation}
where $(a)$ follows from \eqref{equ-p(u,delta) lower bound}. Considering the above we have,
\begin{align*}
&\frac{1}{2}\log{\left(\frac{1+2u^2P}{2+\delta^2P}\right)}P_r(\xi)\\
&\geq \frac{1}{2}\log{\left(\frac{1+2u^2P}{2+o\left(\frac{1}{\ln{N}}\right)P}\right)}\left(1-o\left(e^{-\phi\sqrt{N}}\right)\right)\\
&\geq \frac{1}{2}\log{\left(1+2u^2P\right)}-\frac{1}{2} -o\left(\frac{1}{\log{N}}\right),
\end{align*}
where $0<\phi<0.5$ but bounded away from zero.  
Note that $u(N)$ and $\delta(N)$ determine the search domain of the servers. They are chosen to provide an asymptotic result. Nevertheless, to determine the two servers for transmission the user searches for the two servers with the closest channel coefficients in absolute values which provide the highest computation rate. He does that by sorting $\b{h}$ according to the absolute values, computing the rate for every consecutive pair, and picking the best pair of servers. Asymptotically with $N$, the above analysis lower bounds the rate that can be achieved using this suggested scheme.
\end{IEEEproof}

Theorem \ref{the-expected rate lower bound with P} shows that a scaling law of $O(\log{(P)})$ can be attained, however, due to the choice of only two servers for transmission the scheme does not scale well with $N$. 
 
\begin{figure}[t]
\centering
    \includegraphics[width=0.6\textwidth]{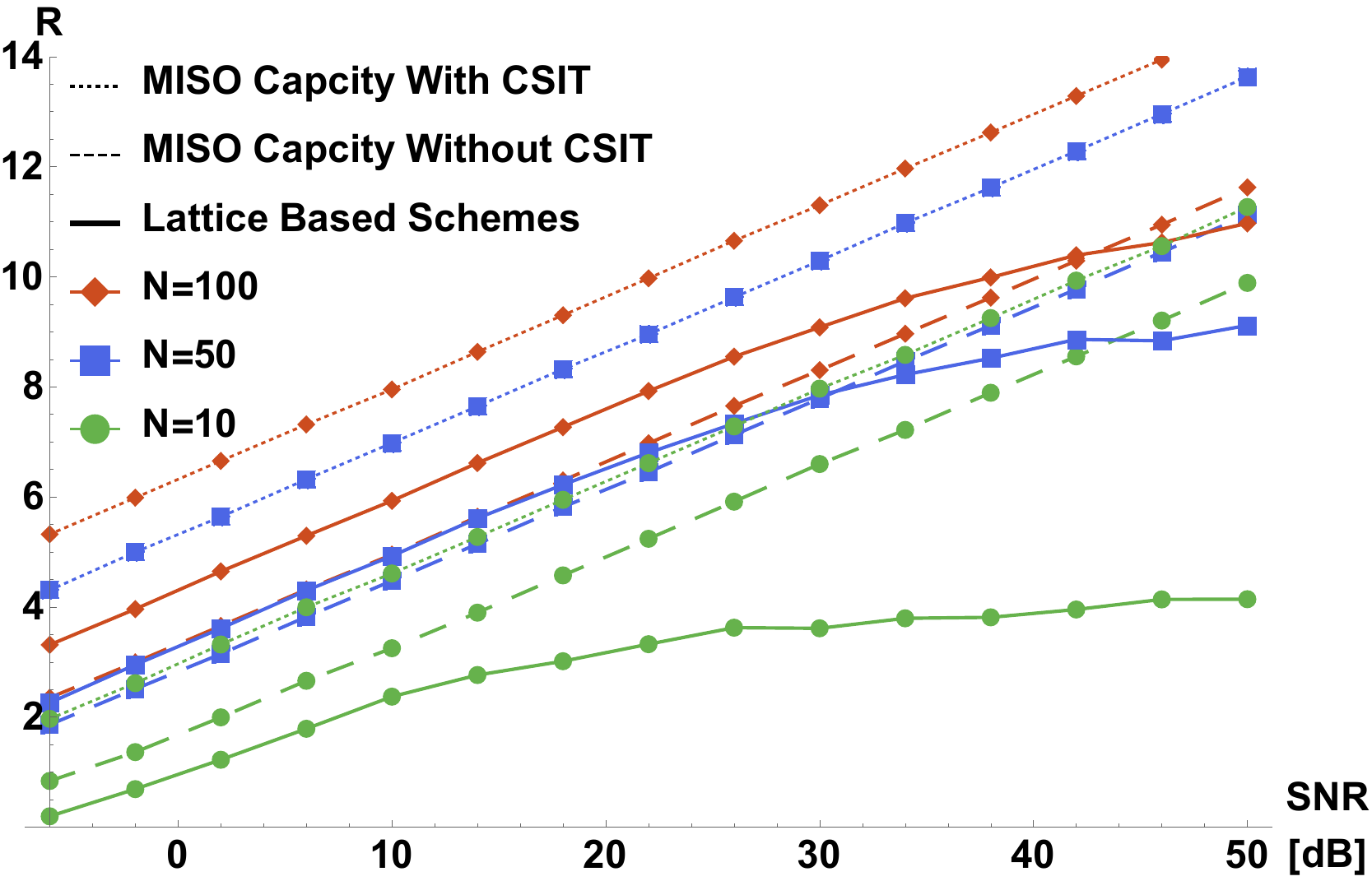}
\caption{The average PIR rate as a function of the SNR for $N=\{10,50,100\}$. The dotted and dashed lines represent the average MISO capacity with and without CSIT as given in \eqref{equ-MISO sum capacity with per-antenna power constraint} and \eqref{equ-ergodic MISO sum capacity with per-antenna power constraint}, respectively. The solid lines are the achievable PIR rate achieved by employing the lattice based scheme given in Theorem \ref{the-retrieval rate for fading MAC} but with subset selection that Algorithm \ref{algo-Greedy Algorithm} provides.}
\label{fig-rate as function of N heuristic}
\end{figure}

To show that the PIR scheme suggested in Theorem \ref{the-retrieval rate for fading MAC} can provide a rate which approaches the capacity as $P$ and $N$ grows, we suggest another scheme for partitioning the sets $\cS_1$ and $\cS_2$. The scheme is based on a greedy heuristic algorithm for solving the number partition problem \cite{karmarkar1982differencing,gent1998analysis} for which we provide numerical results depicted in Figure \ref{fig-rate as function of N heuristic}. The scheme, which is given in Algorithm \ref{algo-Greedy Algorithm}, selects the two sets of servers from only those with positive sign in $\b{h}$ (i.e., from only $N/2$ servers on average) in a greedy way that tries to balance $\tilde{h_1}$ and $\tilde{h_2}$. Figure \ref{fig-rate as function of N heuristic} depicts the PIR rate achieved by employing the scheme given in Theorem \ref{the-retrieval rate for fading MAC} with the partition of $\cS_1$ and $\cS_2$ that Algorithm \ref{algo-Greedy Algorithm} provides. Comparing figures \ref{fig-PIR_rate_fading_MISO_capaciy} and \ref{fig-rate as function of N heuristic} one can observe that the PIR rate is higher than the capacity without CSIT and has a constant gap from the capacity with CSIT. That is, although the servers do not have CSIT, the PIR scheme implicitly generates cooperation as if CSIT exist. The curves diverge for large values of SNR. Where we attribute this to the non-optimality of the algorithm.

\begin{algorithm}[t]
\caption{Greedy Algorithm for $\cS_1$ and $\cS_2$}
\label{algo-Greedy Algorithm}
\hspace*{\algorithmicindent} \textbf{Input:} $\b{h}$ \\
 \hspace*{\algorithmicindent} \textbf{Output:} $\cS_1,\cS_2$
\begin{algorithmic}[1]
\Algphase{Initialization:}
\State $\b{h}^{pos} \gets \text{All positive elements of } \b{h}$
\State $\b{h}^{pos}_s \gets \text{Sort } \b{h}^{pos}$ \text{ from largest to lowest}
\State arrA = $\{\}$; arrB = $\{\}$; sumA = $0$; sumB = $0$;
\Algphase{Main:}
\For {$i =1; \ i\leq \text{length}(\b{h}^{pos}_s); \ i++$} 
 	\If {sumA $<$ sumB}
		\State AppendTo[arrA, i]; 
		\State sumA = sumA + arr[i]; 
	\Else
		\State AppendTo[arrB, i]; 
		\State sumB = sumB + arr[i]; 
	\EndIf
	\EndFor
	\State \Return $(\cS_1,\cS_2)$ \text{ corresponding to } (arrA,arrB)
\end{algorithmic}
\end{algorithm}

\subsection{Scaling Law discussion}
Throughout this paper, to assess the goodness of our results, we made a comparison between the achievable PIR rates and the capacity of the channel with no privacy constraints. Considering the AWGN-MAC with block-fading, we distinguished between the cases in which the channel coefficients are globally known or unknown. The capacities' expressions were given in \eqref{equ-MISO sum capacity with per-antenna power constraint} and \eqref{equ-ergodic MISO sum capacity with per-antenna power constraint}, respectively. The scaling laws of the expected channel capacities, assuming Rayleigh fading, are $O(\log(N^2P))$ and $O(\log(NP))$, respectively\footnote{An upper bound on the expected rates in \eqref{equ-MISO sum capacity with per-antenna power constraint} and \eqref{equ-ergodic MISO sum capacity with per-antenna power constraint} can be obtained easily by using Jensen's inequality.}. Thus, knowing the CSI provides a gain of $2$ to the first case. 

The suggested schemes in theorems \ref{the-expected rate lower bound} and \ref{the-expected rate lower bound with P} show that a scaling law of either $O(\log{(N)})$ or $O(\log{(P)})$ can be guaranteed analytically depending on the partition of servers into $\cS_1$ and $\cS_2$. The first is obtained if one tries to use the statistical properties of $\b{h}$ such that on average there are equal number of positive and negative elements, thus, their sum is approximately equal and grows with $N$. However, the variance of the difference between the sums is unbounded and therefore it does not decrease the denominator in the rate expression, leading to the loss in the scaling with $P$. The second scheme shows that one can always pick two servers with small difference in their channel coefficients magnitude, which decreases the denominator as $N$ grows.

Thus, we conclude that with the suggested sub-optimal schemes the scaling laws of the channel capacity without CSIT can be obtained fully, where for the case of known CSIT we are only twice lower than the optimum. Nevertheless, to promise a rate which scales well simultaneously with $P$ and $N$, one must analyze a scheme which is based on the specific realization of the channel's coefficients. Such a task is complex and may not have a closed form solution.

\section{Conclusion}
In this work, we considered the problem of PIR over an AWGN MAC, with and without fading. The AWGN MAC is a highly non-trivial channel for this problem, with noisy, interfering and possibly asymmetric links. Accordingly, the PIR scheme must take into account the restrictions imposed due to the channel characteristics. We showed that for such AWGN MAC, the PIR scheme and the channel coding scheme should be designed jointly to attain better performance compared to schemes that rely on separation. 

We provided joint privacy-channel coding retrieval schemes for both non-fading and fading channels. The achievable rates were shown to be up to a constant gap from the channel capacity without privacy constraints. Moreover, as the number of servers grows, these rates are asymptotically optimal, as they approach the capacity. Similar behavior was shown for SNR scaling.

%\textcolor{blue}{. As future work, it would be interesting to check whether a universally better joint scheme exists which will outperform separation even for the low SNR regime.}

\appendices

\bibliographystyle{IEEEtran}
\bibliography{../Bibliography_PIR}

\begin{thebibliography}{10}
\providecommand{\url}[1]{#1}
\csname url@samestyle\endcsname
\providecommand{\newblock}{\relax}
\providecommand{\bibinfo}[2]{#2}
\providecommand{\BIBentrySTDinterwordspacing}{\spaceskip=0pt\relax}
\providecommand{\BIBentryALTinterwordstretchfactor}{4}
\providecommand{\BIBentryALTinterwordspacing}{\spaceskip=\fontdimen2\font plus
\BIBentryALTinterwordstretchfactor\fontdimen3\font minus
  \fontdimen4\font\relax}
\providecommand{\BIBforeignlanguage}[2]{{%
\expandafter\ifx\csname l@#1\endcsname\relax
\typeout{** WARNING: IEEEtran.bst: No hyphenation pattern has been}%
\typeout{** loaded for the language `#1'. Using the pattern for}%
\typeout{** the default language instead.}%
\else
\language=\csname l@#1\endcsname
\fi
#2}}
\providecommand{\BIBdecl}{\relax}
\BIBdecl

\bibitem{chor1995private}
B.~Chor, O.~Goldreich, E.~Kushilevitz, and M.~Sudan, ``Private information
  retrieval,'' in \emph{Proceedings of IEEE 36th Annual Foundations of Computer
  Science}.\hskip 1em plus 0.5em minus 0.4em\relax IEEE, 1995, pp. 41--50.

\bibitem{gasarch2004survey}
W.~Gasarch, ``A survey on private information retrieval,'' \emph{Bulletin of
  the EATCS}, vol.~82, no. 72-107, p. 113, 2004.

\bibitem{ostrovsky2007survey}
R.~Ostrovsky and W.~E. Skeith, ``A survey of single-database private
  information retrieval: Techniques and applications,'' in \emph{International
  Workshop on Public Key Cryptography}.\hskip 1em plus 0.5em minus 0.4em\relax
  Springer, 2007, pp. 393--411.

\bibitem{yekhanin2010private}
S.~Yekhanin, ``Private information retrieval,'' in \emph{Locally Decodable
  Codes and Private Information Retrieval Schemes}.\hskip 1em plus 0.5em minus
  0.4em\relax Springer, 2010, pp. 61--74.

\bibitem{sun2017capacity}
H.~Sun and S.~A. Jafar, ``The capacity of private information retrieval,''
  \emph{IEEE Transactions on Information Theory}, vol.~63, no.~7, pp.
  4075--4088, 2017.

\bibitem{sun2017capacityColluding}
------, ``The capacity of robust private information retrieval with colluding
  databases,'' \emph{IEEE Transactions on Information Theory}, vol.~64, no.~4,
  pp. 2361--2370, 2017.

\bibitem{tajeddine2017private}
R.~Tajeddine, O.~W. Gnilke, D.~Karpuk, R.~Freij-Hollanti, C.~Hollanti, and
  S.~El~Rouayheb, ``Private information retrieval schemes for coded data with
  arbitrary collusion patterns,'' in \emph{2017 IEEE International Symposium on
  Information Theory (ISIT)}.\hskip 1em plus 0.5em minus 0.4em\relax IEEE,
  2017, pp. 1908--1912.

\bibitem{banawan2018capacityByzantine}
K.~Banawan and S.~Ulukus, ``The capacity of private information retrieval from
  byzantine and colluding databases,'' \emph{IEEE Transactions on Information
  Theory}, vol.~65, no.~2, pp. 1206--1219, 2018.

\bibitem{sun2018capacity}
H.~Sun and S.~A. Jafar, ``The capacity of symmetric private information
  retrieval,'' \emph{IEEE Transactions on Information Theory}, vol.~65, no.~1,
  pp. 322--329, 2018.

\bibitem{sun2017optimal}
------, ``Optimal download cost of private information retrieval for arbitrary
  message length,'' \emph{IEEE Transactions on Information Forensics and
  Security}, vol.~12, no.~12, pp. 2920--2932, 2017.

\bibitem{tian2019capacity}
C.~Tian, H.~Sun, and J.~Chen, ``Capacity-achieving private information
  retrieval codes with optimal message size and upload cost,'' \emph{IEEE
  Transactions on Information Theory}, vol.~65, no.~11, pp. 7613--7627, 2019.

\bibitem{yang2018private}
H.~Yang, W.~Shin, and J.~Lee, ``Private information retrieval for secure
  distributed storage systems,'' \emph{IEEE Transactions on Information
  Forensics and Security}, vol.~13, no.~12, pp. 2953--2964, 2018.

\bibitem{banawan2020private}
K.~Banawan and S.~Ulukus, ``Private information retrieval through wiretap
  channel {I}{I}: Privacy meets security,'' \emph{IEEE Transactions on
  Information Theory}, 2020.

\bibitem{wei2019capacity}
Y.-P. Wei, K.~Banawan, and S.~Ulukus, ``The capacity of private information
  retrieval with partially known private side information,'' \emph{IEEE
  Transactions on Information Theory}, 2019.

\bibitem{shah2014one}
N.~B. Shah, K.~Rashmi, and K.~Ramchandran, ``One extra bit of download ensures
  perfectly private information retrieval,'' in \emph{2014 IEEE International
  Symposium on Information Theory}.\hskip 1em plus 0.5em minus 0.4em\relax
  IEEE, 2014, pp. 856--860.

\bibitem{chan2015private}
T.~H. Chan, S.-W. Ho, and H.~Yamamoto, ``Private information retrieval for
  coded storage,'' in \emph{2015 IEEE International Symposium on Information
  Theory (ISIT)}.\hskip 1em plus 0.5em minus 0.4em\relax IEEE, 2015, pp.
  2842--2846.

\bibitem{tajeddine2018private}
R.~Tajeddine, O.~W. Gnilke, and S.~El~Rouayheb, ``Private information retrieval
  from {MDS} coded data in distributed storage systems,'' \emph{IEEE
  Transactions on Information Theory}, vol.~64, no.~11, pp. 7081--7093, 2018.

\bibitem{banawan2018capacity}
K.~Banawan and S.~Ulukus, ``The capacity of private information retrieval from
  coded databases,'' \emph{IEEE Transactions on Information Theory}, vol.~64,
  no.~3, pp. 1945--1956, 2018.

\bibitem{zhu2019new}
J.~Zhu, Q.~Yan, C.~Qi, and X.~Tang, ``A new capacity-achieving private
  information retrieval scheme with (almost) optimal file length for coded
  servers,'' \emph{IEEE Transactions on Information Forensics and Security},
  vol.~15, pp. 1248--1260, 2019.

\bibitem{tajeddine2019private}
R.~Tajeddine, A.~Wachter-Zeh, and C.~Hollanti, ``Private information retrieval
  over random linear networks,'' \emph{IEEE Transactions on Information
  Forensics and Security}, vol.~15, pp. 790--799, 2019.

\bibitem{banawan2019noisy}
K.~Banawan and S.~Ulukus, ``Noisy private information retrieval: On
  separability of channel coding and information retrieval,'' \emph{IEEE
  Transactions on Information Theory}, 2019.

\bibitem{nazer2007computation}
B.~Nazer and M.~Gastpar, ``Computation over multiple-access channels,''
  \emph{IEEE Transactions on information theory}, vol.~53, no.~10, pp.
  3498--3516, 2007.

\bibitem{banawan2019asymmetry}
K.~Banawan and S.~Ulukus, ``Asymmetry hurts: Private information retrieval
  under asymmetric traffic constraints,'' \emph{IEEE Transactions on
  Information Theory}, vol.~65, no.~11, pp. 7628--7645, 2019.

\bibitem{zhu2016gaussian}
J.~Zhu and M.~Gastpar, ``Gaussian multiple access via compute-and-forward,''
  \emph{IEEE Transactions on Information Theory}, vol.~63, no.~5, pp.
  2678--2695, 2016.

\bibitem{cover2012elements}
T.~M. Cover and J.~A. Thomas, \emph{Elements of information theory}.\hskip 1em
  plus 0.5em minus 0.4em\relax John Wiley \& Sons, 2012.

\bibitem{vu2011miso}
M.~Vu, ``Miso capacity with per-antenna power constraint,'' \emph{IEEE
  Transactions on Communications}, vol.~59, no.~5, pp. 1268--1274, 2011.

\bibitem{erez2004achieving}
U.~Erez and R.~Zamir, ``Achieving 1/2 log (1+ {SNR}) on the {AWGN} channel with
  lattice encoding and decoding,'' \emph{Information Theory, IEEE Transactions
  on}, vol.~50, no.~10, pp. 2293--2314, 2004.

\bibitem{erez2005capacity}
U.~Erez, S.~Shamai, and R.~Zamir, ``Capacity and lattice strategies for
  canceling known interference,'' \emph{IEEE Transactions on Information
  Theory}, vol.~51, no.~11, pp. 3820--3833, 2005.

\bibitem{nam2008capacity}
W.~Nam, S.-Y. Chung, and Y.~H. Lee, ``Capacity bounds for two-way relay
  channels,'' in \emph{2008 IEEE International Zurich Seminar on
  Communications}.\hskip 1em plus 0.5em minus 0.4em\relax IEEE, 2008, pp.
  144--147.

\bibitem{wilson2010joint}
M.~P. Wilson, K.~Narayanan, H.~D. Pfister, and A.~Sprintson, ``Joint physical
  layer coding and network coding for bidirectional relaying,'' \emph{IEEE
  Transactions on Information Theory}, vol.~56, no.~11, pp. 5641--5654, 2010.

\bibitem{nazer2011compute}
B.~Nazer and M.~Gastpar, ``Compute-and-forward: Harnessing interference through
  structured codes,'' \emph{IEEE Transactions on Information Theory}, vol.~57,
  no.~10, pp. 6463--6486, 2011.

\bibitem{hong2013compute}
S.-N. Hong and G.~Caire, ``Compute-and-forward strategies for cooperative
  distributed antenna systems,'' \emph{IEEE Transactions on Information
  Theory}, vol.~59, no.~9, pp. 5227--5243, 2013.

\bibitem{zhan2014integer}
J.~Zhan, B.~Nazer, U.~Erez, and M.~Gastpar, ``Integer-forcing linear
  receivers,'' \emph{IEEE Transactions on Information Theory}, vol.~60, no.~12,
  pp. 7661--7685, 2014.

\bibitem{ordentlich2014approximate}
O.~Ordentlich, U.~Erez, and B.~Nazer, ``The approximate sum capacity of the
  symmetric gaussian $ k $-user interference channel,'' \emph{IEEE Transactions
  on Information Theory}, vol.~60, no.~6, pp. 3450--3482, 2014.

\bibitem{lim2018joint}
S.~H. Lim, C.~Feng, A.~Pastore, B.~Nazer, and M.~Gastpar, ``A joint typicality
  approach to compute-forward,'' \emph{IEEE Transactions on Information
  Theory}, vol.~64, no.~12, pp. 7657--7685, 2018.

\bibitem{shmuel2020compute}
O.~Shmuel, A.~Cohen, and O.~Gurewitz, ``Compute-and-forward in large relaying
  systems: Limitations and asymptotically optimal scheduling,'' \emph{arXiv
  preprint arXiv:2005.04592}, 2020.

\bibitem{ordentlich2016simple}
O.~Ordentlich and U.~Erez, ``A simple proof for the existence of ``good'' pairs
  of nested lattices,'' \emph{IEEE Transactions on Information Theory},
  vol.~62, no.~8, pp. 4439--4453, 2016.

\bibitem{zamir2014lattice}
R.~Zamir, \emph{Lattice Coding for Signals and Networks: A Structured Coding
  Approach to Quantization, Modulation, and Multiuser Information
  Theory}.\hskip 1em plus 0.5em minus 0.4em\relax Cambridge University Press,
  2014.

\bibitem{cormen2009introduction}
T.~H. Cormen, C.~E. Leiserson, R.~L. Rivest, and C.~Stein, \emph{Introduction
  to algorithms}.\hskip 1em plus 0.5em minus 0.4em\relax MIT press, 2009.

\bibitem{karmarkar1982differencing}
N.~Karmarkar and R.~M. Karp, \emph{The differencing method of set
  partitioning}.\hskip 1em plus 0.5em minus 0.4em\relax Computer Science
  Division (EECS), University of California Berkeley, 1982.

\bibitem{gent1998analysis}
I.~P. Gent and T.~Walsh, ``Analysis of heuristics for number partitioning,''
  \emph{Computational Intelligence}, vol.~14, no.~3, pp. 430--451, 1998.

\end{thebibliography}

\end{document}